\documentclass[%
 onecolumn,
 amsmath,amssymb,
 aps,
]{revtex4-1}
\def\comment#1{}
\usepackage{mathrsfs}
\usepackage{graphicx}
\usepackage{dcolumn}
\usepackage{bm}

\begin{document}
\preprint{APS/123-QED}

\title{Gravitational Radiations from a Spinning Compact Object around \\a Supermassive Kerr Black Hole in Circular Orbit}

\author{Wen-Biao Han}
 \altaffiliation[Also at ]{Physics Department, University of Rome ``La Sapienza," P.le A. Moro 5, 00185 Rome, Italy}
 \email{wenbiao@icra.it}
\affiliation{%
 ICRANet Piazzale della Repubblica, 10-65122, Pescara, Italy 
}%




\date{\today}

\begin{abstract}
The gravitational waves and energy radiations from a spinning
compact object with stellar mass in a circular orbit in the
equatorial plane of a supermassive Kerr black hole are investigated
in this paper. The effect how the spin acts on energy and angular
moment fluxes is discussed in detail. The calculation results
indicate that the spin of small body should be considered in
waveform-template production for the upcoming gravitational wave
detections. It is clear that when the direction of spin axes is the
same as the orbitally angular momentum (``positive'' spin), spin can
decrease the energy fluxes which radiate to infinity. For
antidirection spin (``negative''), the energy fluxes to infinity can
be enlarged. And the relations between fluxes (both infinity and
horizon) and spin look like quadratic functions. From frequency
shift due to spin, we estimate the wave-phase accumulation during
the inspiraling process of the particle. We find that the time of
particle inspiral into the black hole is longer for positive spin
and shorter for negative compared with the nonspinning particle.
Especially, for extreme spin value, the energy radiation near the
horizon of the extreme Kerr black hole is much more than that for
the nonspinning one. And consequently, the maximum binging energy
 of the extreme spinning particle is much larger than that of the nonspinning
 particle.
\begin{description}
\item[PACS numbers]
\end{description}
\end{abstract}

\pacs{Valid PACS appear here}
\maketitle

\section{Introduction}
It is a popular opinion that at the center of our Galaxy and many
galaxies there are supermassive black
holes~\cite{SMBH1}-\cite{SMBH4}. And extreme mass ratio inspiral
(EMRI): a compact stellar-mass object (black hole, neutron star or
white dwarf) spiraling into the supermassive black hole in galactic
nuclei is one of the most important sources of gravitational waves
for NASA's Laser Interferometer Space Antenna (LISA)~\cite{Lisa}.
People estimated that LISA should see about 50 such events of EMRIs
at distances out to redshifts $z\gtrsim
1~$\cite{event1}\cite{event2}\cite{event3}, and the signal-to-noise
ratios of these sources are around 10-100~\cite{Finn}. Furthermore,
the gravitational signals from EMRIs can also offer very useful
information to study the mass distribution of galactic nuclei and
the basic physics near the horizon of black holes, such as the
no-hair theorem.

For the motives mentioned above, we should have enough accurate and
numerous waveform templates to match the upcoming observational data
in the future. A very universal method to calculate the
gravitational wave templates is the black hole perturbation
technology developed by
Teukolsky~\cite{Teukolsky1}\cite{Teukolsky74}. The Teukolsky
equation determines the (linearized) evolution of a perturbation to
the Kerr spacetime. There are many published papers investigating
gravitational radiations and orbital evolution. The Teukolsky
equation is solved in these papers by post-Newtonian expansion
(\cite{pn1}\cite{pn2}and references inside) and numerical simulation
(\cite{Hughes1}-\cite{Tagoshi3} and references inside).

Usually, the study of EMRIs did not containe spins of the small mass
particles. Because for the spinning particle in the Kerr spacetime,
the dynamical equation is non-integrable, different from the
spinless one. The motion of spinning particles is much more
difficult to work out. Only a few papers considered gravitational
waves from the spinning particles for some special orbits by using
the Teukolsky equation. Mino, Shibata and Tanaka first investigated
gravitational waves induced by a spinning particle falling into a
rotating black hole\cite{spin1}. Then Tanaka et al. researched the
case of particle with small spin in circular orbits around a Kerr
black hole\cite{spin2}.

On the other hand, when the spin is measured by terms of $\mu M$,
the spin parameter is
\begin{align}
S\leq\frac{\mu^2}{\mu M}=\frac{\mu}{M}\ll 1,
\end{align}
where $\mu, M$ are the mass of the small object and central black
hole respectively. People think that the spins of small objects are
so tiny that they can be ignored without obvious influence.

But small spin can produce considerable accumulated effect when we
 observe the secular evolution of EMRIs, and can also make the phase of
gravitational waves different from the spinless case. The importance
of spin was emphasized by Burko who considered a spinning compact
object in a quasicircular, planar orbit around a Schwarzschild black
hole \cite{Burko1}\cite{Burko2}. The effect of spin is at most
marginally relevant for signal detection \cite{Barack}. And we will
see that the results of this paper can prove the spin of small body
cannot be ignored.

Furthermore, for the bigger value of spin, perhaps the first order
perturbation method is not accurate enough, but the result can also
offer some useful information on intermediate-mass-ratio inspirals
(IMRIs). Especially, when spin achieves extreme value ($S\sim 1$),
some very interesting phenomena
 appear. For example, in the Schwarzschild or Kerr background,
researchers found chaotic behavior of extreme spinning
particles\cite{spin3}-\cite{spin6}. Even without chaos, extreme spin
 can also produce some distinct orbital characteristics \cite{spin6}.
Obviously, gravitational signals of chaotic orbits would be much
different from those of regular ones. References \cite{spin7} and
\cite{Hanphd} by using quadruple radiation formula calculated
gravitational waveforms of chaotic systems.

In the present paper, as the beginning of future research, we study
the gravitation radiation of spinning particles in circular orbits
on an equatorial plane by calculating the Teukolsky equation. In
Ref.~\cite{spin2}, the authors used the post-Newtonian expansion and
assumed very small spin to keep the circular orbits. Differing from
them, we calculate the Teukolsky equation numerically and make the
particles be restricted on equatorial plane exactly. So we can get
the results in a strong field where the small objects are near the
black holes ($r\sim M$). The signals from such a strong-field area
are
 easier for the detectors to catch. And also we can study
gravitational radiation and orbital evolution when the value of spin
becomes much bigger.

As mentioned above, the strong spin of particle can produce chaotic
orbits in some parameter range. But the circular orbits studied in
this paper, can be determined completely and off course are
nonchaotic. This ensures the stability of numerical results. In the
future, we will discuss the gravitational waves and radiation
reaction of chaotic systems in another paper.

This paper is organized as follows. In the next section, we first
use the Papapetrou equation to deduce the orbital parameters of the
spinning particle in the equatorially circular orbit. Second, we
introduce the Teukolsky-Sasaki-Nakamura formalisms briefly and our
numerical method. In Sec. IV we present our results in detail.
Finally, our conclusion and discussion are given in the last
section.

We use units $G=c=1$ and the metric signature $(-,+,+,+)$. Distance
and time are measured by the central black hole mass $M$, and the
energy of the particle is measured by $\mu$, the mass of itself. We
measure the orbital angular momentum and spin of the particle in
unit of $\mu M$.

\section{Circular orbit of a spinning particle}
The equations of motion of a spinning test particle in a curved
spacetime were given first by Papapetrou \cite{papapetrou}, and then
reformulated more clearly form by Dixon \cite{Dixon2}, In our units
 the can be written as
\begin{align}
\frac{d x^\mu}{d\tau}&=v^\mu, \label{spin1}\\
\frac{d p^\mu}{d\tau}&=-(\frac{1}{2}R^{*\mu}_{~\nu\alpha\beta}
p^\alpha S^\beta+\Gamma^\mu_{\nu\alpha}p^{\alpha})v^\nu, \label{spin2}\\
\frac{d S^\mu}{d\tau}&=-(p^\mu R^{*}_{\nu\alpha\beta\gamma}S^\alpha
p^\beta S^\gamma+\Gamma^\mu_{\nu\alpha}S^{\alpha})v^\nu.
\label{spin3}
\end{align}
Where, $v^\mu$ is the four-velocity, $p^\mu$ is the linear momentum
and $S^\mu$ the spin vector. The latter two must satisfy
\begin{align}
p^\nu p_\nu=-\mu^2, ~ S^\nu S_\nu=S^2,
\end{align}
where $S$ represents spin magnitude. We define
\begin{align}
R^{*}_{\mu\nu\alpha\beta}=\frac{1}{2}R^{~\rho\sigma}_{\mu\nu}\epsilon_{\rho\sigma\alpha\beta},
\end{align}
and $\epsilon^{\rho\sigma\alpha\beta}$ is the Levi-Civita tensor.

Because of spin, the motion of particle does not follow the
geodesic, then the $p^\mu$ is no longer parallel to $v^\mu$. If
following Dixon, choosing the rest frame of the particle¡¯s center
of mass, we can get one of the spin supplementary conditions,
\begin{align}
p^\mu S_{\mu}=0,
\end{align}
and the relation between four-velocity and linear momentum
\begin{align}
v^\mu=u^\mu+\frac{2S^{\mu\nu}R_{\nu\alpha\beta\gamma}u^\alpha
S^{\beta\gamma}}{4+R_{\alpha\beta\gamma\delta}S^{\alpha\beta}S^{\gamma\delta}},\label{v_p}
\end{align}
where $u^\nu=p^\nu/\mu$ and spin tensor is
\begin{align}
S^{\mu\nu}=\epsilon^{\mu\nu\alpha\beta}S_\alpha u_{\beta}.
\end{align}

As with the spinless particle, there are two conservations, energy
and total angular momentum which includes orbit and spin angular
momentum
\begin{align}
E&=-p_t+\frac{1}{2}g_{t\mu,\nu}S^{\mu\nu},\label{energy}\\
J_z&=p_\phi-\frac{1}{2}g_{\phi\mu,\nu}S^{\mu\nu}.\label{momentum}
\end{align}
Another constant, the Carter constant $Q$ for the nonspinning
particle, does not exit for a spinning one anymore.

Usually, solving Eq.~(\ref{spin1}-\ref{spin3}) is difficult. By
assuming the spinning particle moving in circular orbit on the
equatorial plane of a Kerr black hole, the problem can be simplified
greatly. In this assumption, the only nonzero component of spin must
be $s\equiv S^\theta=-S/r$. Then, using the condition of circular
orbit $dp^r/d\tau=0$ , combining Eq.(\ref{spin2}) and the relation
(\ref{v_p}), after a long calculation, we get the angular frequency
\begin{equation}
\Omega_\phi \equiv d \phi/d t=\frac{\sum_{i=0}^2 a_i s^i \mp
\sqrt{\sum_{i=0}^{4}b_i s^i}}{\sum_{i=0}^2 c_i
s^i},\label{frequency}
\end{equation}
where $s^i$ presents the $i$th power of $s$. The upper sign refers
to prograde and the lower to retrograde orbits and
\begin{align}
a_0&=2Mar,~ a_1=-(3r^2+6a^2)M, ~a_2=(4M^2a+3Mar),~b_0=4Mr^5, \\
b_1&=-12Mar^4,~
b_2=13M^2r^4,~b_3=-6M^2ar^3,~ b_4=(9a^2-8Mr)M^2r^2,\\
c_0&=2(Ma^2r-r^4),~ c_1=-6Ma(a^2+r^2),~ c_2=(4M^2a^2+6Ma^2r+2Mr^3).
\end{align}
\comment{
\begin{align}
\Omega_\phi \equiv d \phi/d t
\\=\frac{2Mar-(3Mr^2+6Ma^2)s+(4M^2a+3Mar)s^2 \mp
\sqrt{4Mr^5-12Mar^4s+13M^2r^4s^2-6M^2ar^3s^3+(9a^2-8Mr)M^2r^2s^4}}{2(Ma^2r-r^4)-6Ma(a^2+r^2)s+(4M^2a^2+6Ma^2r+2Mr^3)s^2},
\end{align}}
This result coincides with the approximative expression Eq. (4.26)
in Ref.(\ref{spin2}) when $s\ll 1$. Obviously, Eq. (\ref{frequency})
reduces to the angular frequency of the spinless particle if the
spin $s=0$,
\begin{align}
\Omega_\phi \equiv d \phi/d t=\pm\frac{M^{1/2}}{r^{3/2}\pm
aM^{1/2}},
\end{align}
 As mentioned before, because of the spin, the four-velocity is not
parallel with linear momentum as before. From Eq.~(\ref{v_p}),
relation between them is
\begin{eqnarray}
v^t&=&u^t\left[1-\frac{M(3a^2+r^2)s^2}{\mu^2 r^3}\right]+\frac{3Ma(a^2+r^2)}{\mu^2 r^3}s^2u^\phi,\\
v^\phi&=&u^\phi\left[1+\frac{M(3a^2+2r^2)s^2}{\mu^2
r^3}\right]-\frac{3Mas^2u^t}{\mu^2 r^3}.
\end{eqnarray}
We can clearly see that the difference between $u^\mu$ and $v^\mu$
is the $s^2$ order. And from the energy (\ref{energy}) and total
angular momentum (\ref{momentum}), we can get,
\begin{align}
u_t&=-\frac{E}{\mu}+\left(\frac{z_1E/\mu-XJ_z}{\mu+z_1 s}\right)s,\label{cmomentum0}\\
u_\phi &=\frac{J_z}{\mu}+\left(\frac{z_2 J_z/\mu+YE}{\mu-z_2
s}\right)s,\label{cmomentum3}
\end{align}
where,
\begin{align}
X &=\frac{\frac{M}{r^2}}{\mu-\frac{Ma}{r^2}s},\\
Y &=\frac{(r-\frac{Ma^2}{r^2})}{\mu+\frac{Ma}{r^2}s},\\
z_1 &=
\frac{M}{r^2}\left[a-\frac{(r-\frac{Ma^2}{r^2})s}{\mu-\frac{Ma}{r^2}s}\right],
\\
z_2 &=
\frac{M}{r^2}\left[a+\frac{(r-\frac{Ma^2}{r^2})s}{\mu+\frac{Ma}{r^2}s}\right].
\end{align}
Furthermore, the effective 4-velocity components
\begin{align}
u^t &=u_0^t+\left\{[\frac{a^2}{r^2}-\frac{(r^2+a^2)^2}{\Delta
r^2}]\mathcal{X}+\frac{a}{r^2}(1-\frac{r^2+a^2}{\Delta})\mathcal{Y}\right\}s,\label{momentum0}\\
u^\phi
&=u_0^\phi+\left\{[1-\frac{a^2}{\Delta}]\mathcal{Y}+\frac{a^2}{r^2}(1-\frac{r^2+a^2}{\Delta})\mathcal{X}\right\}s,
\label{momentum3}
\end{align}
where $u_0^t, u_0^\phi$ are the components without spin, and
$\mathcal{X}, \mathcal{Y}$ just are the coefficients before $s$ in
Eq.(\ref{cmomentum0},\ref{cmomentum3}).

For confirming a circular orbit, first the Kerr parameter $a$ and
spin magnitude $S$ are fixed. Then we give a radius $r_0$, from
which
\begin{equation}
\frac{\partial U(r)}{\partial r}\Bigr|_{r_0}=0,\quad
E=U(r_0),\label{circleeq}
\end{equation}
can determine the total angular momentum $J_z$ and energy $E$. Where
$U(r)$ is the so-called effective potential,
\begin{equation}
U(J_z,a,S,r)=\frac{B}{A}J_z+\sqrt{(\frac{B^2}{A^2}-\frac{C}{A})J_z^2+\frac{\mu^2
Z^2}{A}},
\end{equation}
and the related coefficients are
\begin{align}
Z &=\sqrt{\Delta}(1-\frac{Ms^2}{r\mu^2})^2/[1-(\frac{Mas}{r^2\mu})^2],\\
A &=[\Delta+(r^2+a^2)\frac{2M}{r}](\mu-z_2s)^2-(1-\frac{2M}{r})(\mu+z_1s)^2Y^2s^2+\frac{4Ma}{r}ZYs,\\
B &=(1-\frac{2M}{r})(\mu+z_1s)^2Ys-[\Delta+(r^2+a^2)\frac{2M}{r}](\mu-z_2s)^2Xs-\frac{2Ma}{r}Z(1+XYs^2),\\
C
&=[\Delta+(r^2+a^2)\frac{2M}{r}](\mu-z_2s)^2X^2s^2-(1-\frac{2M}{r})(\mu+z_1s)^2+\frac{4Ma}{r}ZXs
\end{align}
Finally the four-momentum components $p^t,p^\phi$ can be calculated
from Eq.(\ref{momentum0},\ref{momentum3}).

Now, we deduce the energy-momentum tensor of a spinning particle in
curved spacetime. From Dixon's classical literature \cite{Dixon}, we
can write down the energy-momentum tensor,
\begin{align}\nonumber
T^{\alpha\beta}(x)&=\mu\int{d\tau\left\{\frac{\delta^{(4)}(x-z(\tau))}{\sqrt{-g}}u^{(\alpha}(x,\tau)v^{\beta)}(x,\tau)-\nabla_{\gamma}\left(S^{\gamma
(\alpha}(x,\tau)v^{\beta)}(x,\tau)\frac{\delta^{(4)}(x-z(\tau))}{\sqrt{-g}}\right)\right\}}\\\nonumber
&=\mu\int{d\tau\left\{\frac{\delta^{(4)}(x-z(\tau))}{\sqrt{-g}}[u^{(\alpha}v^{\beta)}-\Gamma_{\gamma\delta}^{(\alpha}v^{\beta)}S^{\gamma\delta}+\Gamma_{\gamma\delta}^{(\alpha}S^{\beta)\gamma}v^{\delta}]-\frac{\partial_{\gamma}[S^{\gamma
(\alpha}v^{\beta)}\delta^{(4)}(x-z(\tau))]1}{\sqrt{-g}}\right\}}\\
&\equiv\mu\int{d\tau\left\{\frac{\delta^{(4)}(x-z(\tau))}{\sqrt{-g}}U^{\alpha\beta}-\frac{[\partial_{\gamma}V^{\gamma
\alpha\beta}\delta^{(4)}(x-z(\tau))]}{\sqrt{-g}}\right\}}\equiv\mu\mathcal{T}^{\alpha\beta}.
\label{energy-momentum tensor1}
\end{align}
where $z(\tau)$ is the world line of the spinning particle. For
convenience, we write $U^{\alpha\beta}=u^{(\alpha}
v^{\beta)}+U^{\alpha\beta}_{(s)}s$. Considering there are only two
independent components of the spin tensor in the circular
orbit\cite{Hanphd},
\begin{equation}
S^{tr}=-S^{rt}=su_{\phi},\quad S^{r\phi}=-S^{\phi r}=su_{t},
\end{equation}
the nonzero components of $U^{\alpha\beta}_{(s)}$ are
\begin{align}\nonumber
U^{tt}_{(s)}&=\Gamma^t_{rt}u_\phi u^t+\Gamma^t_{r\phi}u_\phi
u^\phi,\\ \nonumber
U^{t\phi}_{(s)}&=\frac{1}{2}(\Gamma^\phi_{rt}u_\phi
u^t+\Gamma^\phi_{r\phi}u_\phi u^\phi-\Gamma^t_{rt}u_t
u^t-\Gamma^t_{r\phi}u_t u^\phi),\\
\nonumber U^{rr}_{(s)}&=\Gamma^r_{\phi t}u_t
u^t+\Gamma^r_{\phi\phi}u_t u^\phi-\Gamma^r_{tt}u_\phi
u^t-\Gamma^r_{t\phi}u_\phi u^\phi,\\
U^{\phi\phi}_{(s)}&=-\Gamma^\phi_{rt}u_t u^t+\Gamma^\phi_{r\phi}u_t
u^\phi.\label{e-m1}
\end{align}
\comment{
\begin{align}\nonumber
U^{tt}&=(\Gamma^t_{rt}u_\phi v^t+\Gamma^t_{r\phi}u_\phi v^\phi)s,\\
\nonumber U^{t\phi}&=\frac{1}{2}(\Gamma^\phi_{rt}u_\phi
v^t+\Gamma^\phi_{r\phi}u_\phi v^\phi-\Gamma^t_{rt}u_t
v^t-\Gamma^t_{r\phi}u_t v^\phi)s,\\ \nonumber
U^{rr}&=(\Gamma^r_{\phi t}u_t v^t+\Gamma^r_{\phi\phi}u_t
v^\phi-\Gamma^r_{tt}u_\phi v^t-\Gamma^r_{t\phi}u_\phi v^\phi)s,\\
\nonumber U^{\phi\phi}&=-(\Gamma^\phi_{rt}u_t
v^t+\Gamma^\phi_{r\phi}u_t v^\phi)s.
\end{align}
}
 And the nonzero components of $V^{\gamma\alpha\beta}$ are
\begin{align}\nonumber
V^{trt}&=\frac{1}{2}u_{\phi}v^ts,\quad
V^{tr\phi}=\frac{1}{2}u_{\phi}v^{\phi}s;\\\nonumber V^{\phi
rt}&=-\frac{1}{2}u_{t}v^ts,\quad V^{\phi
r\phi}=-\frac{1}{2}u_{t}v^{\phi}s;\\
V^{rt\phi}&=\frac{1}{2}(u_{t}v^{t}-u_{\phi}v^{\phi})s.\label{e-m2}
\end{align}
\comment{
\begin{align}\nonumber
V^{trt}_{(s)}&=\frac{1}{2}u_{\phi}u^t,\quad
V^{tr\phi}_{(s)}=\frac{1}{2}u_{\phi}u^{\phi};\\\nonumber V^{\phi
rt}_{(s)}&=-\frac{1}{2}u_{t}u^t,\quad V^{\phi
r\phi}_{(s)}=-\frac{1}{2}u_{t}u^{\phi};\\
V^{rt\phi}_{(s)}&=\frac{1}{2}(u_{t}u^{t}-u_{\phi}u^{\phi}).
\end{align}
} \comment{Furthermore(delete),
\begin{equation}
T^{\alpha\beta}(t,r,\theta,\phi)=\frac{\mu}{\Sigma \sin{\theta}
\dot{t}}\left[u^{(\alpha}v^{\beta)}-\nabla_{\gamma}(S^{\gamma(\alpha}v^{\beta)})\right]\delta(r-r(t))\delta(\theta-\theta(t))\delta(\phi-\phi(t)),
\label{energy-momentum tensor2}
\end{equation}}
Now, we project this energy-momentum tensor onto the Newman-Penrose
null tetrad\cite{Chandrabook},
\begin{align}
n_\alpha &=\frac{1}{2}\left(\frac{\Delta}{\Sigma},1,0,-\frac{a\Delta\sin^2\theta}{\Sigma}\right),\\
\bar{m}_{\alpha}
&=\frac{\rho}{\sqrt{2}}\left(ia\sin\theta,0,\Sigma,-i(r^2+a^2)\sin\theta\right).
\label{Newman-Penrose}
\end{align}
It can be,
\begin{equation}
T_{nn}=n_{\alpha}n_{\beta}T^{\alpha\beta},\quad
T_{n\bar{m}}=n_{\alpha}\bar{m}_{\beta}T^{\alpha\beta},\quad
T_{\bar{m}\bar{m}}=\bar{m}_{\alpha}\bar{m}_{\beta}T^{\alpha\beta},
\end{equation}
where $\rho=(r-ia\cos\theta)^{-1}$. So,
\begin{equation}
T_{ab}=\mu C_{ab}
\end{equation}
\comment{\begin{equation} T_{ab}=\frac{\mu
C_{ab}}{\Sigma\sin\theta\dot{t}}\delta(r-r(t))\delta(\theta-\theta(t))\delta(\phi-\phi(t))
\end{equation}}
where $C_{ab}$ can be expanded as
\begin{align}
C_{nn}&=\frac{1}{4}[\frac{\Delta^2}{\Sigma^2}\mathcal{T}^{00}+\mathcal{T}^{11}+\frac{a^2\Delta^2\sin^4\theta}{\Sigma^2}\mathcal{T}^{33}+2\frac{\Delta}{\Sigma}\mathcal{T}^{01}-2\frac{a\Delta^2\sin^2\theta}{\Sigma^2}\mathcal{T}^{03}-2\frac{a\Delta\sin^2\theta}{\Sigma}\mathcal{T}^{13}],\label{Cnn}\\\nonumber
C_{n\bar{m}}&=\frac{\rho}{2\sqrt{2}}[\frac{ia\Delta\sin\theta}{\Sigma}(\mathcal{T}^{00}+(a^2+r^2)\sin^2\theta
\mathcal{T}^{33})+ia\sin\theta\mathcal{T}^{01}+\Delta\mathcal{T}^{02}-\\
&\quad\frac{i\Delta\sin\theta}{\Sigma}(r^2+a^2+a^2\sin^2\theta)\mathcal{T}^{03}+\Sigma\mathcal{T}^{12}-i(r^2+a^2)\sin\theta\mathcal{T}^{13}-a\Delta\sin^2\theta\mathcal{T}^{23}
],\label{Cnm}\\\nonumber
C_{\bar{m}\bar{m}}&=\frac{\rho^2}{2}[-a^2\sin^2\theta\mathcal{T}^{00}+\Sigma^2\mathcal{T}^{22}-(r^2+a^2)^2\sin^2\theta\mathcal{T}^{33}+2ia\Sigma\sin\theta\mathcal{T}^{02}+\\
&\quad
2a(r^2+a^2)\sin^2\theta\mathcal{T}^{03}-2i\Sigma(r^2+a^2)\sin\theta\mathcal{T}^{23}].\label{Cmm}
\end{align}
\comment{The definition of $\mathcal{T}^{\mu\nu}$ is
\begin{equation}
\mathcal{T}^{\mu\nu}\equiv
u^{(\mu}v^{\nu)}-\nabla_{\gamma}(S^{\gamma(\mu}v^{\nu)})
\end{equation}}
For circular orbits on the equatorial plane, the above equations can
be greatly simplified.

\section{Gravitational radiation and radiation reaction}
\subsection{The Teukolsky-Sasaki-Nakamura formalism} In this paper,
we use the Teukolsky equation to calculate gravitational radiation
and radiation reaction.The Teukolsky formalism considers
perturbation on the Weyl curvature scale $\psi_4$, which can be
decomposed in the frequency domain~\cite{Teukolsky1},
\begin{align}
\psi_4=\rho^4\int^{+\infty}_{-\infty}{d\omega\sum_{lm}{R_{lm\omega}(r)_{~-2}S^{a\omega}_{lm}(\theta)e^{im\phi}e^{-i\omega
t}}},
\end{align}
where $\rho=-1/(r-ia\cos{\theta})$. The function $R_{lm\omega}(r)$
satisfied the radial Teukolsky equation
\begin{align}
\Delta^2\frac{d}{dr}\left(\frac{1}{\Delta}\frac{d
R_{lm\omega}}{dr}\right)-V(r)R_{lm\omega}=-\mathscr{T}_{lm\omega}(r),\label{Teukolsky}
\end{align}
where $\mathscr{T}_{lm\omega}(r)$ is the source term, which will be
given below, and the potential is
\begin{align}
V(r)=-\frac{K^2+4i(r-M)K}{\Delta}+8i\omega r+\lambda,
\end{align}
where $K=(r^2+a^2)\omega-ma, ~\lambda=E_{lm}+a^2\omega^2-2a m w-2$.
The spin-weighted angular function $_{-2}S^{a\omega}_{lm}(\theta)$
obeys the following equation,
\begin{align}
\frac{1}{\sin{\theta}}\frac{d}{d\theta}\left(\sin{\theta
\frac{d_{~-2}S^{a\omega}_{lm}}{d\theta}}\right)+\left[(a\omega)^2\cos^2{\theta}+4a\omega\cos{\theta}-\left(\frac{m^2-4m\cos\theta+4}{\sin^2\theta}\right)+E_{lm}\right]
~_{-2}S^{a\omega}_{lm}=0.
\end{align}
The radial Teukolsky Eq.(\ref{Teukolsky}) has the general solution
\begin{align}
R_{lm\omega}(r)=\frac{R^{\infty}_{lm\omega}(r)}{2i\omega
B^{in}_{lm\omega}D^{\infty}_{lm\omega}}\int^{r}_{r_+}{dr'\frac{R^H_{lm\omega}(r')\mathscr{T}_{lm\omega}(r')}{\Delta(r')^2}}+\frac{R^{H}_{lm\omega}(r)}{2i\omega
B^{in}_{lm\omega}D^{\infty}_{lm\omega}}\int^{\infty}_{r}{dr'\frac{R^\infty_{lm\omega}(r')\mathscr{T}_{lm\omega}(r')}{\Delta(r')^2}},\label{solutiont}
\end{align}
where the $R^{\infty}_{lm\omega}(r)$ and $R^{H}_{lm\omega}(r)$ are
two independent solutions of the homogeneous Teukolsky equation.
They are chosen to be the purely ingoing wave at the horizon and
purely outgoing wave at infinity respectively,
\begin{align}\nonumber
R^{H}_{lm\omega}&=B^{hole}_{lm\omega}\Delta^2 e^{-ipr*},\quad
r\rightarrow r_+\\
R^{H}_{lm\omega}&=B^{out}_{lm\omega}r^3 e^{i\omega
r*}+r^{-1}B^{in}_{lm\omega}r e^{-i\omega r*},\quad r\rightarrow
\infty;
\end{align}
and
\begin{align}\nonumber
R^{\infty}_{lm\omega}&=D^{\rm{out}}_{lm\omega} e^{ip r*}+\Delta^2
D^{in}_{lm\omega}r e^{-ip r*},\quad
r\rightarrow r_+\\
R^{\infty}_{lm\omega}&=r^3 D^{\infty}_{lm\omega} e^{-i\omega
r*},\quad r\rightarrow \infty,
\end{align}
where $k=\omega-ma/2Mr_+$ and $r*$ is the ``tortoise coordinate".
The solution (\ref{solutiont}) must be purely ingoing at horizon and
purely outgoing at infinity. That is,
\begin{align}
R_{lm\omega}(r\rightarrow \infty)&=Z^{H}_{lm\omega}r^3 e^{i\omega
r*},\\
R_{lm\omega}(r\rightarrow r_+)&=Z^{\infty}_{lm\omega}\Delta^2 e^{-ip
r*}.
\end{align}
The complex amplitudes $Z^{H,\infty}_{lm\omega}$ are defined as
\begin{align}\nonumber
Z^{H}_{lm\omega}&=\frac{1}{2i\omega
B^{in}_{lm\omega}}\int^{r}_{r_+}{dr'\frac{R^H_{lm\omega}(r')\mathscr{T}_{lm\omega}(r')}{\Delta(r')^2}},\\
Z^{\infty}_{lm\omega}&=\frac{B^{hole}_{lm\omega}}{2i\omega
B^{in}_{lm\omega}D^{\infty}_{lm\omega}}\int^{\infty}_{r}{dr'\frac{R^\infty_{lm\omega}(r')\mathscr{T}_{lm\omega}(r')}{\Delta(r')^2}}.\label{Z1}
\end{align}

The particle in this paper is in circular orbit on the equatorial
plane, thus the particle's motion is described only as the harmonic
of the frequency $\Omega_\phi$ (\ref{frequency}), and defined
\begin{align}
\omega_m=m\Omega_\phi.
\end{align}
Then $Z^{H,\infty}_{lm\omega}$ are decomposed as
\begin{align}\nonumber
Z^{H}_{lm\omega}&=Z^{H}_{lm}\delta(\omega-\omega_m),\\
Z^{\infty}_{lm\omega}&=Z^{\infty}_{lm}\delta(\omega-\omega_m).\label{Z2}
\end{align}
And the amplitudes $Z^{H,\infty}_{lm}$ fully determine the energy
and angular momentum fluxes of gravitational radiations.

\subsection{Energy, angular momentum fluxes and waveforms}

From the Weyl scalar $\psi_4$, we can extract the two polarization
components $h_+,~h_{\times}$ of the transverse-traceless metric
perturbation at $r\rightarrow\infty$,
\begin{equation}
\psi_4(r\rightarrow\infty)\rightarrow\frac{1}{2}(\ddot{h}_+-i\ddot{h}_\times).
\end{equation}
The energy flux in gravitational waves, from the Isaacson
stress-energy tensor\cite{Isaacson} is
\begin{equation}
\left(\frac{dE}{dAdt}\right)_{r\rightarrow\infty}^{\textrm{rad}}=\frac{1}{16\pi}\left\langle\left(\frac{\partial
h_+}{\partial t}\right)^2+\left(\frac{\partial h_\times}{\partial
t}\right)^2\right\rangle,
\end{equation}
then we can get the fluxes at infinity as follow
\begin{align}\nonumber
\left(\frac{dE}{dt}\right)_{r\rightarrow\infty}^{\textrm{rad}}&=\sum_{lm}{\frac{|Z_{lm}^H|^2}{4\pi
\omega_m^2}},\\
\left(\frac{dJ_z}{dt}\right)_{r\rightarrow\infty}^{\textrm{rad}}&=\sum_{lm}{\frac{m|Z_{lm}^H|^2}{4\pi
\omega_m^3}}.\label{flux8}
\end{align}

The energy and angular momentum fluxes at the horizon can be
calculated by the below formula\cite{Teukolsky74},
\begin{align}\nonumber
\left(\frac{dE}{dt}\right)_{r\rightarrow
r_+}^{\textrm{rad}}&=\sum_{lm}{\alpha_{lm}\frac{|Z_{lm}^\infty|^2}{4\pi
\omega_m^2}},\\
\left(\frac{dJ_z}{dt}\right)_{r\rightarrow
r_+}^{\textrm{rad}}&=\sum_{lm}{\alpha_{lm}\frac{m|Z_{lm}^\infty|^2}{4\pi
\omega_m^3}}.\label{fluxh}
\end{align}
Where the coefficient $\alpha_{lm}$ is given as
\begin{equation}
\alpha_{lm}=\frac{256(2Mr_+)p(p^2+4\epsilon^2)(p^2+16\epsilon^2)\omega^3_{m}}{|C_{lm}|^2},
\end{equation}
with $\epsilon=\sqrt{M^2-a^2}/4Mr_+$, and
\begin{align}\nonumber
|C_{lm}|^2=&[(\lambda+2)^2+4am\omega_m-4a^2\omega_m^2](\lambda^2+36ma\omega_m-36a^2\omega_m^2)\\
&+(2\lambda+3)(96a^2\omega^2_m-48ma\omega_m)144\omega^2_m(M^2-a^2).
\end{align}
The total change in energy and angular momentum of the compact
object is
\begin{equation}
\dot{E}=-(\dot{E}^H+\dot{E}^\infty),\quad
\dot{J_z}=-(\dot{J}_z^H+\dot{J}_z^\infty).\label{energyloss}
\end{equation}

From the energy loss and angular momentum loss of radiation, we can
calculate the radiation reaction. There are two key points we can
use to write down the orbit modification in an easy way: one is that
circular orbits remain circular \cite{Kennefick}-\cite{Mino}; the
other one is that $S$ remains constant at first order
\cite{Glampedakis1} under adiabatic radiation reaction. The averaged
rate of change of the radius is
\begin{equation}
\dot{r}=\{\dot{E}-[\frac{B}{A}+\frac{(\frac{B^2}{A^2}-\frac{C}{A})J_z}{E-\frac{B}{A}J_z}]\dot{J}_z\}/\{(\frac{B}{A})'J_z+\frac{(\frac{B^2}{A^2}-\frac{C}{A})'J_z^2+\mu^2(\frac{Z^2}{A})'}{2(E-\frac{B}{A}J_z)}\}.\label{radiationreaction}
\end{equation}
It is obviously that spin can contribute to $\dot{r}$ in the first
order. Finally, the gravitational waveform can be calculated by
\begin{equation}
h_+(\theta,\phi,t)-ih_\times(\theta,\phi,t)=\sum_{lm}{\frac{1}{\omega_m^2}Z^H_{lm}\,_{-2}S^{a\omega_m}_{lm}(\theta)e^{i(m\phi-\omega
t)}}.\label{waveform}
\end{equation}
\subsection{The Sasaki-Nakamura equation}
As mentioned in the above two subsections, for getting
$Z^{H,\infty}_{lm}$, we should integrate the homogenous version of
Eq.(\ref{Teukolsky}). But there is a difficulty when one numerically
integrates Eq.(\ref{Teukolsky}) due to the long-range nature of the
potential $V(r)$ in (\ref{Teukolsky}). In order to solve this
problem, Sasaki and Nakamura developed the Sasaki-Nakamura function
$X(r)$, governed by a short-ranged potential, to replace the
Teukolsky function $R(r)$ \cite{Sasaki}. The Sasaki-Nakamura
equation reads as
\begin{align}
\frac{d^2
X_{lm\omega}}{dr*^2}-F(r)\frac{dX_{lm\omega}}{dr*}-U(r)X_{lm\omega}=0.\label{s-keq}
\end{align}
The functions $F(r),~U(r)$ can be found in Ref.\cite{Sasaki}. The
Sasaki-Nakamura equation also admits two asymptotic solutions,
\begin{align}
X^{H}_{lm\omega}&=e^{-ipr*},\quad r\rightarrow r_+,\\
X^{H}_{lm\omega}&=A^{\rm{out}}_{lm\omega}\bar{P}(r)e^{i\omega
r*}+A^{\rm{in}}_{lm\omega}P(r)e^{-i\omega r*},\quad r\rightarrow
\infty;
\end{align}
and
\begin{align}
X^{\infty}_{lm\omega}&=C^{\rm{out}}_{lm\omega}\bar{P}(r)e^{ip
r*}+C^{\rm{in}}_{lm\omega}P(r)e^{-ip r*},\quad r\rightarrow r_+,\\
X^{\infty}_{lm\omega}&=\bar{P}(r)e^{-i\omega r*},\quad r\rightarrow
\infty,
\end{align}
where $P(r),~\bar{P}(r)$ can be found in Refs.\cite{Dolan},
\cite{Hughes2}.

The solution of Eq.(\ref{s-keq}) is transformed to the solution of
the Teukolsky equation by
\begin{align}
R^{H,\infty}_{lm\omega}=\frac{1}{\eta}\left[\left(\alpha+\frac{\beta_{,r}}{\Delta}\right)\frac{\Delta
X^{H,\infty}_{lm\omega}}{\sqrt{r^2+a^2}}-\frac{\beta}{\Delta}\frac{d}{dr}\frac{\Delta
X^{H,\infty}_{lm\omega}}{\sqrt{r^2+a^2}}\right].
\end{align}
The relations between the coefficients of the Sasaki-Nakamura
function and the Teukolsky function are
\begin{align}
B^{\rm{in}}_{lm\omega}=-\frac{A^{\rm{in}}_{lm\omega}}{4\omega^2},~B^{\rm{hole}}_{lm\omega}=\frac{1}{d_{lm\omega}},~D^{\infty}_{lm\omega}=-\frac{4\omega^2}{c_0},
\end{align}
and the functions $\alpha,~\beta, ~\eta,$ and $d_{lm\omega}$ are
given clearly in Ref.\cite{Sasaki} or in Appendix B of
\cite{Hughes1}.

\subsection{The source term}
Now, we follow Ref.\cite{Hughes1} to calculate the source term which
is the left part of the Teukolsky equation (\ref{Teukolsky}). For
the circular orbit on the equatorial plane($\theta=\pi/2$),
\begin{equation}
\mathscr{T}_{lm\omega}(r)=\int{dt \Delta^2
\{(A_{nn0}+A_{n\bar{m}0}+A_{\bar{m}\bar{m}0})\delta(r-r_0)+\partial_r[(A_{n\bar{m}1}+A_{\bar{m}\bar{m}1})\delta(r-r_0)]+\partial_r^2[A_{\bar{m}\bar{m}2}\delta(r-r_0)]}\}.\label{source}
\end{equation}
All functions in the above equation are given as below:
\begin{align}
A_{nn0}&=-\frac{2\mu\rho^{-3}\bar{\rho}^{-1}C_{nn}}{\Sigma\Delta^2\dot{t}}[L_1^{\dag}L_2^{\dag}\mathcal
{S}+2ia\rho L_2^{\dag}\mathcal{S}],\\
A_{n\bar{m}0}&=-\frac{2\mu\sqrt{2}\rho^{-3}C_{n\bar{m}}}{\Sigma\Delta\dot{t}}\left[\left(\frac{iK}{\Delta}-\rho-\bar{\rho}\right)L_2^{\dag}\mathcal{S}+i\left(\frac{iK}{\Delta}+\rho+\bar{\rho}\right)a\mathcal{S}(\rho-\bar{\rho})\right],\\
A_{\bar{m}\bar{m}0}&=\frac{\mu\mathcal{S}\rho^{-3}\bar{\rho}C_{\bar{m}\bar{m}}}{\Sigma\dot{t}}\left[\frac{K^2}{\Delta^2}+2i\rho\frac{K}{\Delta}+i\partial_r\left(\frac{K}{\Delta}\right)\right],\\
A_{n\bar{m}1}&=-\frac{2\mu\sqrt{2}\rho^{-3}C_{n\bar{m}}}{\Sigma\Delta\dot{t}}[L_2^{\dag}\mathcal{S}+ia\rho(\rho-\bar{\rho})\mathcal{S}],\\
A_{\bar{m}\bar{m}1}&=\frac{2\mu\mathcal{S}\rho{-3}\bar{\rho}C_{\bar{m}\bar{m}}}{\Sigma\dot{t}}\left(\rho-\frac{iK}{\Delta}\right),\\
A_{\bar{m}\bar{m}2}&=-\frac{\mu\mathcal{S}\rho^{-3}\bar{\rho}}{\Sigma\dot{t}}C_{\bar{m}\bar{m}},
\end{align}
where $\mathcal{S}$ just is shorthand of
$_{-2}\mathcal{S}_{lm}^{a\omega}[\theta(t)]$. The projected
energy-momentum tensor components are
\begin{align}
C_{ab}=U_{ab}+i\omega V^t_{\; ab}-imV^\phi_{\; ab}+V^r_{\;
ab}\frac{\partial}{\partial r},\label{spinC}
\end{align}
where $ab$ represents $nn,\,n\bar{m},\,\bar{m}\bar{m}$, and
$U_{ab},~\,V^\gamma_{\;ab}$ can been calculated from
$U^{\alpha\beta},~V^{\gamma\alpha\beta}$ using
Eq.~(\ref{Cnn})-(\ref{Cmm}). We can find that for the spinning
particle there are three additional terms in the above equation
compared to the nonspinning one. And
\begin{align}
L_2^{\dag}\mathcal{S}&=a\omega\sin\theta\mathcal{S}-\sum^{\infty}_{k=l_\textrm{min}}{{b_k[(k-1)(k+2)]^{1/2}}_{-1}Y_{km}(\theta)},\\
L_1^{\dag}L_2^{\dag}\mathcal{S}&=\sum^{\infty}_{k=l_\textrm{min}}{{b_k[(k-1)k(k+1)(k+2)]^{1/2}}_0Y_{km}(\theta)+2a\omega\sin\theta
L_2^{\dag}\mathcal{S}-(a\omega\sin\theta)^2\mathcal{S}}.
\end{align}
Where $b_l$ represents the coefficients of the eigenvector of
$\mathcal{S}$, $_0Y_{lm}(\theta)$ and $_{-1}Y_{lm}(\theta)$ are
spin-weighted spherical harmonics. Of course, we should choose
$\theta=\pi/2$ in our case.

Now, we can directly write down $Z_{lm}^H,Z_{lm}^\infty$. Taking the
source term (\ref{source}) into Eq.(\ref{Z1}), then using
Eq.(\ref{Z2}), we have
\begin{align}
Z_{lm}^H &=\frac{\pi}{i\omega_m
B_{lm}^{\rm{in}}}I_{lm}^H(r_0),\\
Z_{lm}^{\infty} &=-\frac{\pi}{4i\omega_m^3d_{lm\omega}
B_{lm}^{\rm{in}}}I_{lm}^\infty(r_0), \label{sourcefinal}
\end{align}
where,
\begin{equation}
I^{H,\infty}_{lm}(r_0)=R_{lm\omega}^{H,\infty}(r_0)[A_{nn0}+A_{n\bar{m}0}+A_{\bar{m}\bar{m}0}]-\frac{d
R_{lm\omega}^{H,\infty}}{dr}\Bigr|_{r_0}[A_{n\bar{m}1}+A_{\bar{m}\bar{m}1}]+\frac{d^2
R_{lm\omega}^{H,\infty}}{dr^2}\Bigr|_{r_0}A_{\bar{m}\bar{m}2}.
\end{equation}

\comment{ It is clearly that
\begin{align}
U^{\alpha\beta}&=u^\alpha u^\beta+U^{\alpha\beta}_{(s)}s+O(s^2),\\
V^{\gamma\alpha\beta}&=V^{\gamma\alpha\beta}_{(s)}s+O(s^3).
\end{align}
}
\subsection{Numerical method}
Now, we introduce simply our numerical method for calculating the
gravitational radiation of the spinning particle inspiring Kerr
black hole. First, we need determine the orbital parameters of the
particle. Setting Kerr parameter $a$, spin magnitude $S$ and orbit
radius, we solve Eq.(\ref{circleeq}) to get energy and angular
momentum of the spinning particle. Then based on the discussions in
Sec. II, it is easy to get the angular frequency, four-velocity and
linear momentum.

Second, we numerically integrate the Sasaki-Nakamura equation to get
$Z_{lm}^H,Z_{lm}^\infty$ in the loop on $l=2,3,4,\cdots$ and
$m=-l,\cdots,l$. But considering the symmetry of $Z_{lm}^{H,\infty}$
about harmonics $m$, we only need a loop on $m$ from $1$ to $l$.
Usually (based on the experience of many calculations done before)
it is enough for obtaining satisfying accuracy to loop on $2$ up to
$8$.

Then from Eqs.(\ref{flux8}), (\ref{fluxh}) and (\ref{energyloss}),
we can calculate the energy and angular momentum loss due to the
gravitational radiation. Finally, the orbit change of radiation
reaction and waveform are calculated by
Eqs.(\ref{radiationreaction}) and (\ref{waveform}).

For checking the validation of our code, we compare our results with
the post-Newtonian expansion at weak field without spin. They match
very well. An an example, we give the comparison of the energy
fluxes at infinity from $a=0$ to $1$ while $r=25M$ in Fig. 1. In
addition, we compare our results with the published data of Fujita
and Tagoshi \cite{Tagoshi1} when spin is zero, and find they have
good agreement too. After this check, we think our numerical results
are reliable.

\begin{figure}[!h]
\begin{center}
\includegraphics[width=4.5in]{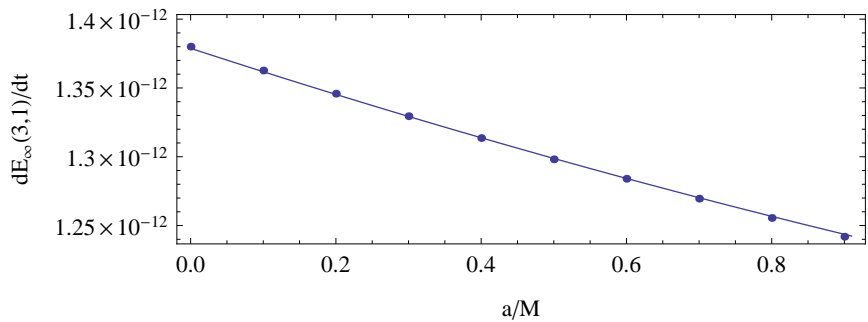}
\caption{Comparison of the energy flux at infinity for orbit at
$r=25M$ as a function of Kerr parameter $a$ for $l=3,~m=1$.
Agreement between the numerical and post-Newtonian fluxes is good.}
\label{check}
\end{center}
\end{figure}

\section{Numerical results}
In this section, we present the numerical results of our
calculations. We focus on how spin influences the gravitational
waves.
\subsection{Orbital frequency with spin}
\begin{figure}[!h]
\begin{center}
\includegraphics[width=4.5in]{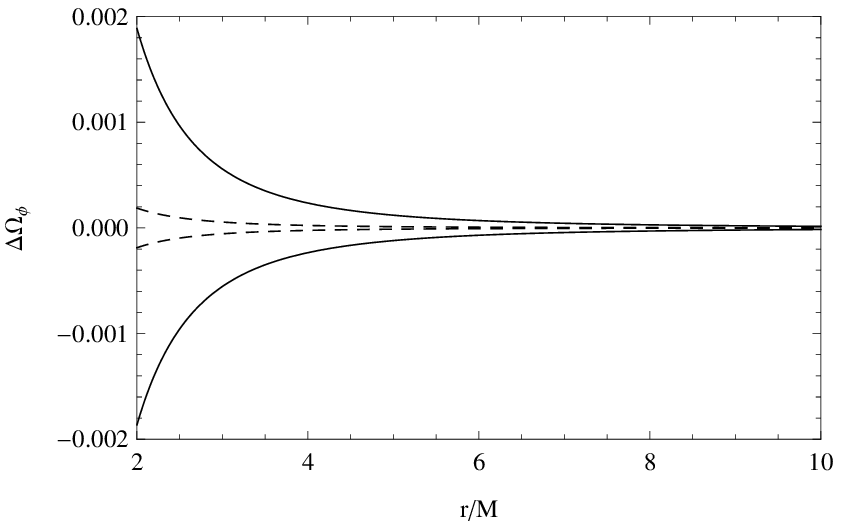}
\caption{The shift of the orbital frequencies between spin and
nonspinning particles as a function of radius $r$ while $a=0$. From
the top down, the spin magnitude $S$ is
$-10^{-2},~-10^{-3},~10^{-3},~10^{-2}$ respectively.}
\label{frequencysa0}
\end{center}
\end{figure}

\begin{figure}[!h]
\begin{center}
\includegraphics[width=4.5in]{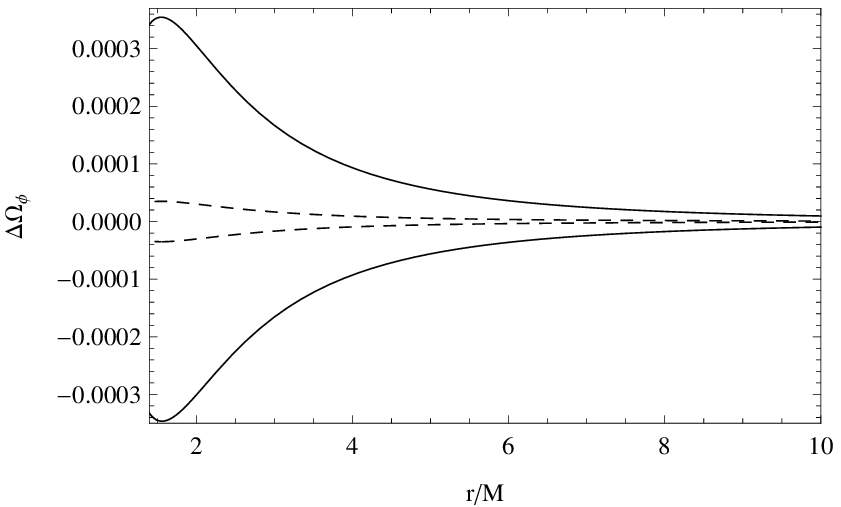}
\caption{The shift of the orbital frequencies between spin and
nonspinning particles as a function of radius $r$ while $a=0.996$.
From the top dowm, the spin magnitude $S$ is
$-10^{-2},~-10^{-3},~10^{-3},~10^{-2}$ respectively.}
\label{frequencysap}
\end{center}
\end{figure}

In Figs.(\ref{frequencysa0}) and (\ref{frequencysap}), we show the
shifts: $\Delta\Omega_\phi=\Omega_\phi(S\neq0)-\Omega_\phi(S=0)$ as
a function of the orbital radius $r$ while $a=0$ and $0.996$
respectively. It is clear that when the direction of spin inverses,
the orbital frequency is bigger than the nonspinning particle. This
is because the total angular momentum is composed of the orbital
angular momentum and the spin angular momentum of the particle. If
spin has the same direction as the total angular momentum, the
orbital angular momentum would decrease, whereas the orbital angular
momentum must be added to counteract the spin. In addition, as the
rotating angular momentum of the black hole increase, both the
orbital angular momentum and the influence of spin on the angular
momentum decrease. Usually, the absolute value of
$\Delta\Omega_\phi$ increases as the orbital radius $r$ decreases.
But near the horizon of the Kerr black hole, perhaps there are some
exceptions, for example, in Fig.(\ref{frequencysap}), the
$S=10^{-2}$ one.

And in Tables \ref{table1} and \ref{table2}, we list the detailed
values of the frequency shifts at several orbital radiuses. It can
be found that the influence of physical spin on the orbital
frequency, as well as the frequency of gravitational waves, is small
and can be negligible. But during the long evolution time of EMRIs
$\sim M^2/\mu$, the accumulatively phasic error without considering
the spin of the inspiraling object is considerable. The phasic shift
due to spin should be calculated by an integration
 $\Delta\phi=\int^{t(r_{\rm{in}})}_{t(r_{\rm{out}})}{\Delta\Omega_\phi(r(t))dt}$.
 This integration is difficult to calculate out, but we can do a
 simple estimate. For $\mu/M\sim10^{-5}$, the average frequency
 shift is about $10^{-7}$ (see Table \ref{table2}) of the extreme
 Kerr black hole. During highly relativistic regime, inspiraling about $\sim M/\mu$
 cycles, and the typical frequency $\Omega_\phi \sim 10^{-2}$, the dephasing of spinning and nonspinning
 particles is
 \begin{align}
 \Delta\phi\sim\Delta\Omega_\phi\frac{M}{\mu}\frac{2\pi}{\Omega_\phi}\sim 2\pi.
 \end{align}
We can find that though the wave-frequency shift of spin is very
small, during the evolution period in a highly relativistic regime,
it can produce considerable (about a 1 cycle) phase error. From the
above estimate, we think the role of spin in the EMRIs should be
calculated to obtain the template of gravitational waveform
accurately.
\begin{table}
\caption{The orbitally angular frequency shift $\Delta\Omega_\phi$
due to the spin for
 $a=0$.}\label{table1}
\begin{tabular}{c|c c c c c c}
\hline \hline
$S$&$-10^{-3}$&$-10^{-4}$&$-10^{-5}$&$10^{-5}$&$10^{-4}$&$10^{-3}$\\
\hline $r=10$&$1.50008\times 10^{-6}$&$1.50001\times
10^{-7}$&$1.50000\times 10^{-8}$&$-1.50000\times
10^{-8}$&$-1.49999\times 10^{-7}$&$-1.49992\times 10^{-6}$\\
$r=8$&$2.92991\times 10^{-6}$&$2.92971\times 10^{-7}$&$2.92969\times
10^{-8}$&$-2.92969\times
10^{-8}$&$-2.92966\times 10^{-7}$&$-2.92946\times 10^{-6}$\\
$r=6$&$6.94527\times 10^{-6}$&$6.94453\times 10^{-7}$&$6.94445\times
10^{-8}$&$-6.94444\times
10^{-8}$&$-6.94436\times 10^{-7}$&$-6.94362\times 10^{-6}$\\
$r=4$&$2.34426\times 10^{-5}$&$2.34380\times 10^{-6}$&$2.34376\times
10^{-7}$&$-2.34374\times
10^{-7}$&$-2.34370\times 10^{-6}$&$-2.34324\times 10^{-5}$\\
$r=2$&$1.87616\times 10^{-4}$&$1.87512\times 10^{-5}$&$1.87501\times
10^{-6}$&$-1.87499\times 10^{-6}$&$-1.87488\times
10^{-5}$&$-1.87384\times 10^{-4}$\\
 \hline \hline
\end{tabular}
\end{table}

\begin{table}
\caption{The orbitally angular frequency shift $\Delta\Omega_\phi$
due to the spin for
 $a=0.996$.}\label{table2}
\begin{tabular}{c|c c c c c c}
\hline \hline
$S$&$-10^{-3}$&$-10^{-4}$&$-10^{-5}$&$10^{-5}$&$10^{-4}$&$10^{-3}$\\
\hline $r=10$&$9.65822\times 10^{-7}$&$9.65768\times
10^{-8}$&$9.65762\times 10^{-9}$&$-9.65761\times
10^{-9}$&$-9.65756\times 10^{-8}$&$-9.65702\times 10^{-7}$\\
$r=8$&$1.74151\times 10^{-6}$&$1.74137\times 10^{-7}$&$1.74136\times
10^{-8}$&$-1.74135\times
10^{-8}$&$-1.74134\times 10^{-7}$&$-1.74120\times 10^{-6}$\\
$r=6$&$3.61475\times 10^{-6}$&$3.61431\times 10^{-7}$&$3.61426\times
10^{-8}$&$-3.61425\times
10^{-8}$&$-3.61421\times 10^{-7}$&$-3.61377\times 10^{-6}$\\
$r=4$&$9.30694\times 10^{-6}$&$9.30480\times 10^{-7}$&$9.30459\times
10^{-8}$&$-9.30454\times
10^{-8}$&$-9.30433\times 10^{-7}$&$-9.30219\times 10^{-6}$\\
$r=2$&$3.03510\times 10^{-5}$&$3.03302\times 10^{-6}$&$3.03281\times
10^{-7}$&$-3.03276\times 10^{-7}$&$-3.03255\times
10^{-6}$&$-3.03047\times 10^{-5}$\\
$r=1.2$&$2.55531\times 10^{-5}$&$2.55116\times
10^{-6}$&$2.55075\times 10^{-7}$&$-2.55065\times
10^{-7}$&$-2.55024\times
10^{-6}$&$-2.54610\times 10^{-5}$\\
 \hline \hline
\end{tabular}
\end{table}

\comment{
\begin{figure}[!h]
\begin{center}
\includegraphics[width=4.5in]{frequency_s_am.eps}
\caption{The shifty of the orbital frequencies between spin and
non-spinning particles as a function of radius $r$ while $a=-0.996$.
From up to below, $S$ are $-10^{-2},~-10^{-3},~10^{-3},~10^{-2}$
respectively.} \label{frequencysam}
\end{center}
\end{figure}
}

\subsection{Energy fluxes vs spin}
\begin{figure}[!h]
\begin{center}
\includegraphics[width=2.5in]{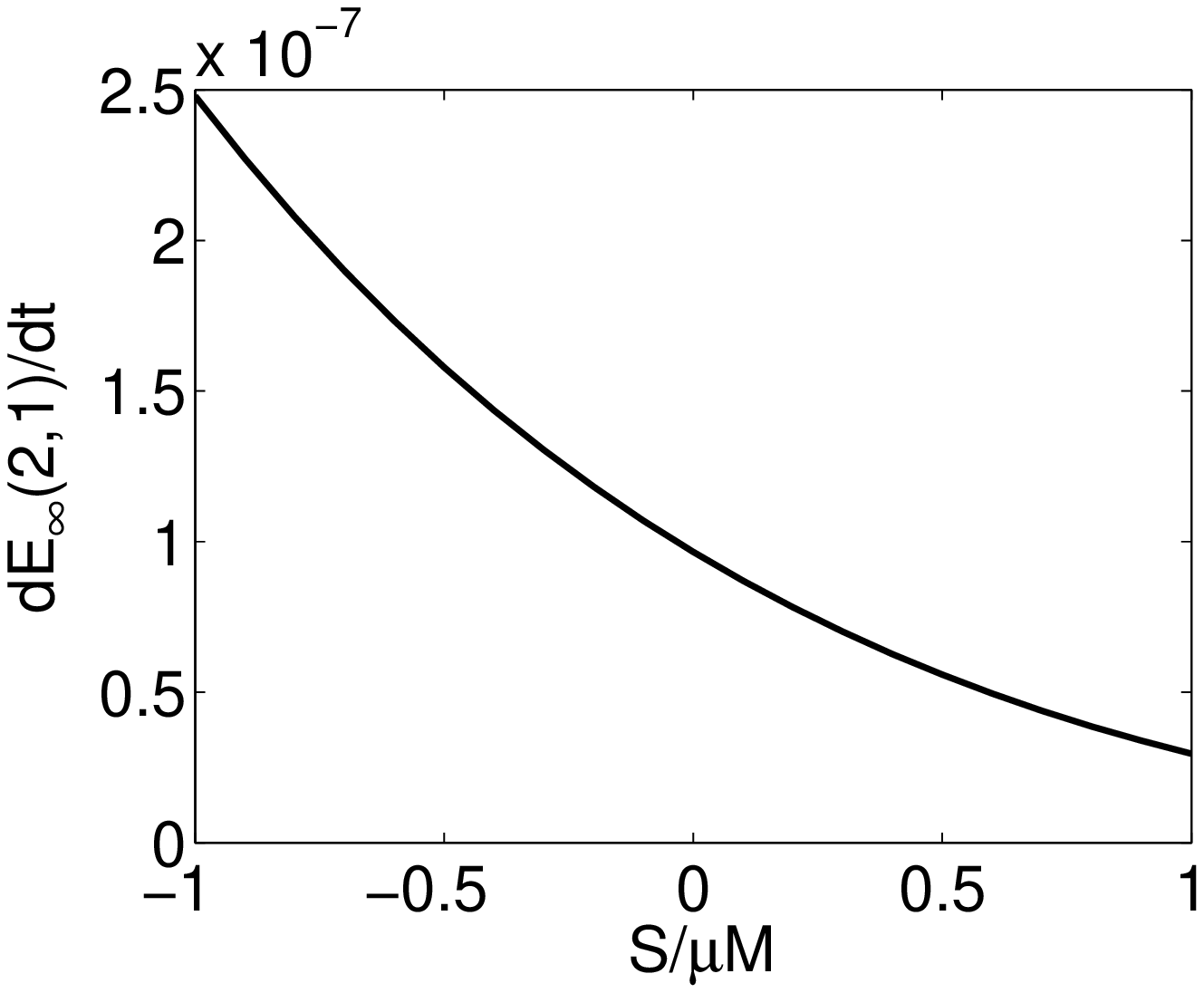}
\includegraphics[width=2.5in]{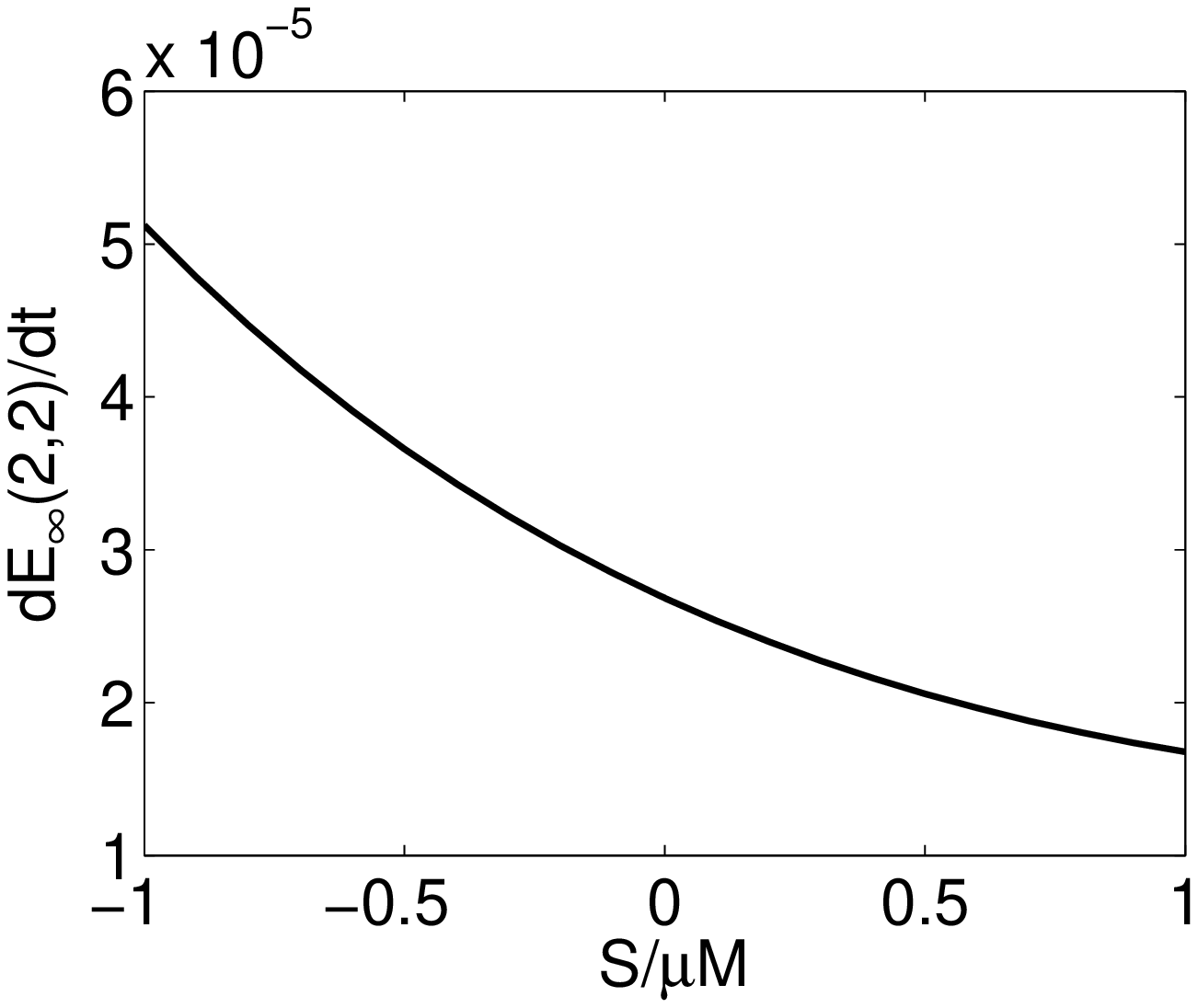}
\includegraphics[width=2.5in]{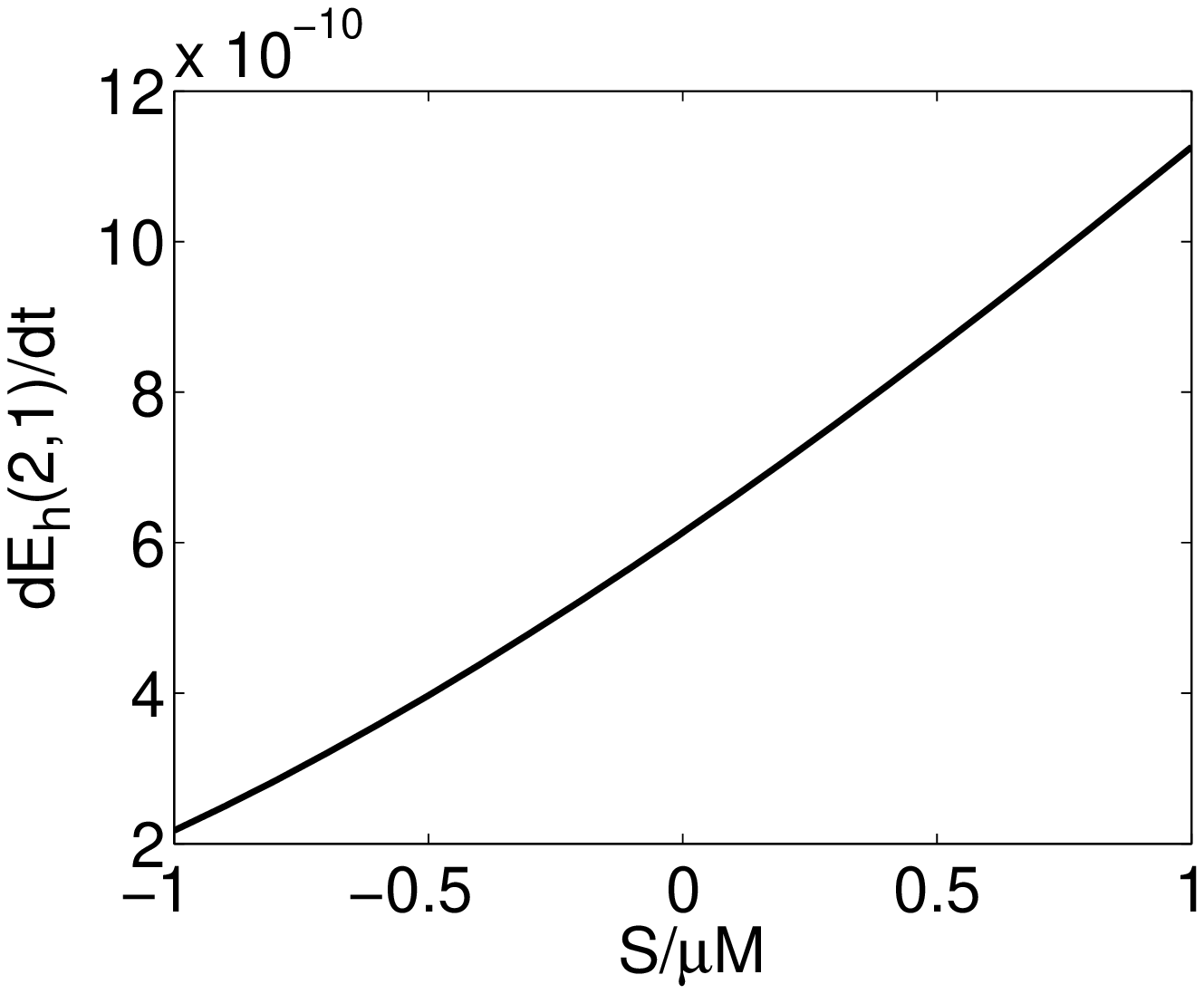}
\includegraphics[width=2.5in]{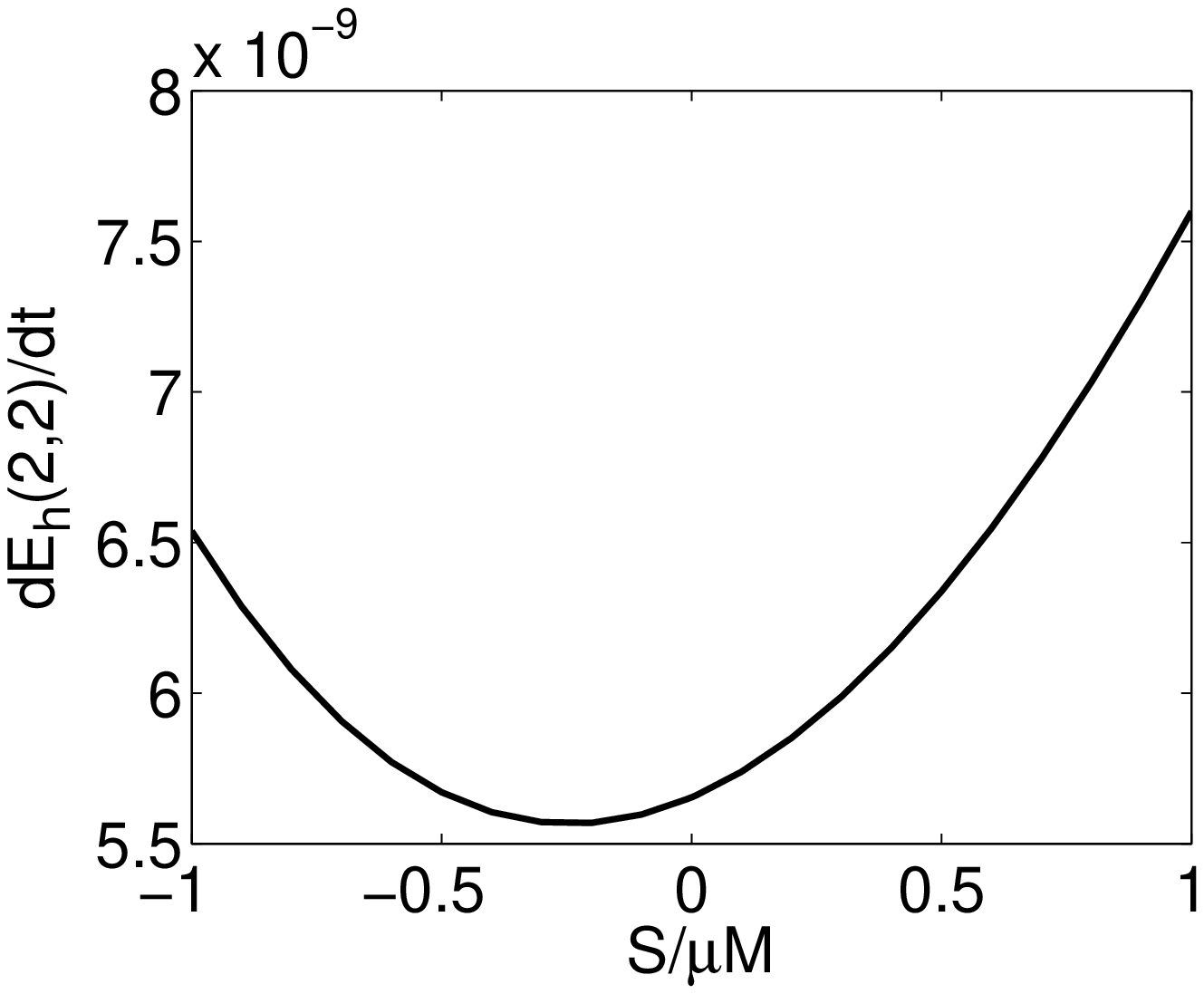}
\caption{The energy fluxes of $l=2$ modes at infinity and the
horizon as a function of the spin magnitude $S$. Where the Kerr
parameter $a=0$ and the orbital radius $r=10M$.} \label{fluxlma0}
\end{center}
\end{figure}

\begin{figure}[!h]
\begin{center}
\includegraphics[width=2.5in]{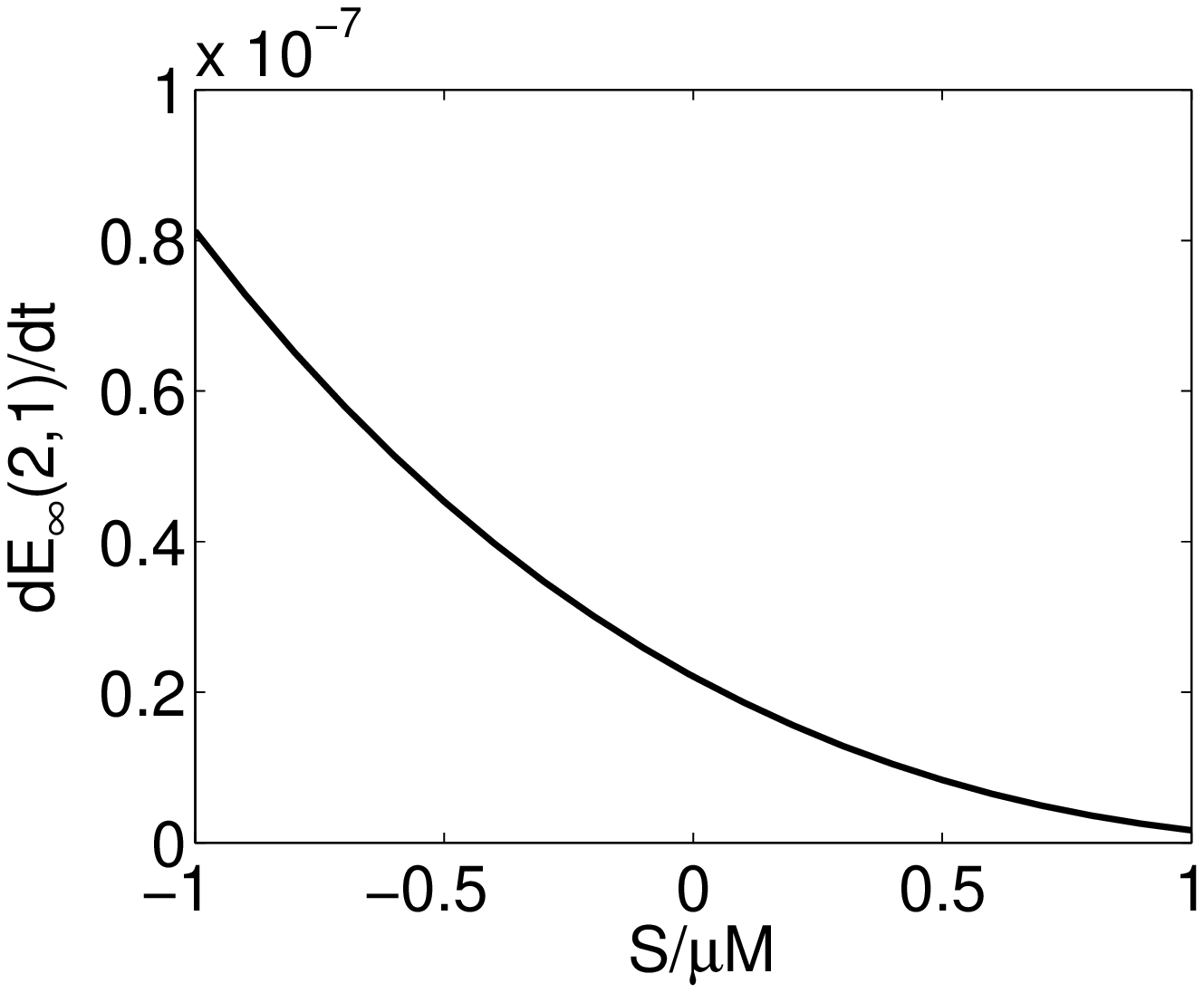}
\includegraphics[width=2.5in]{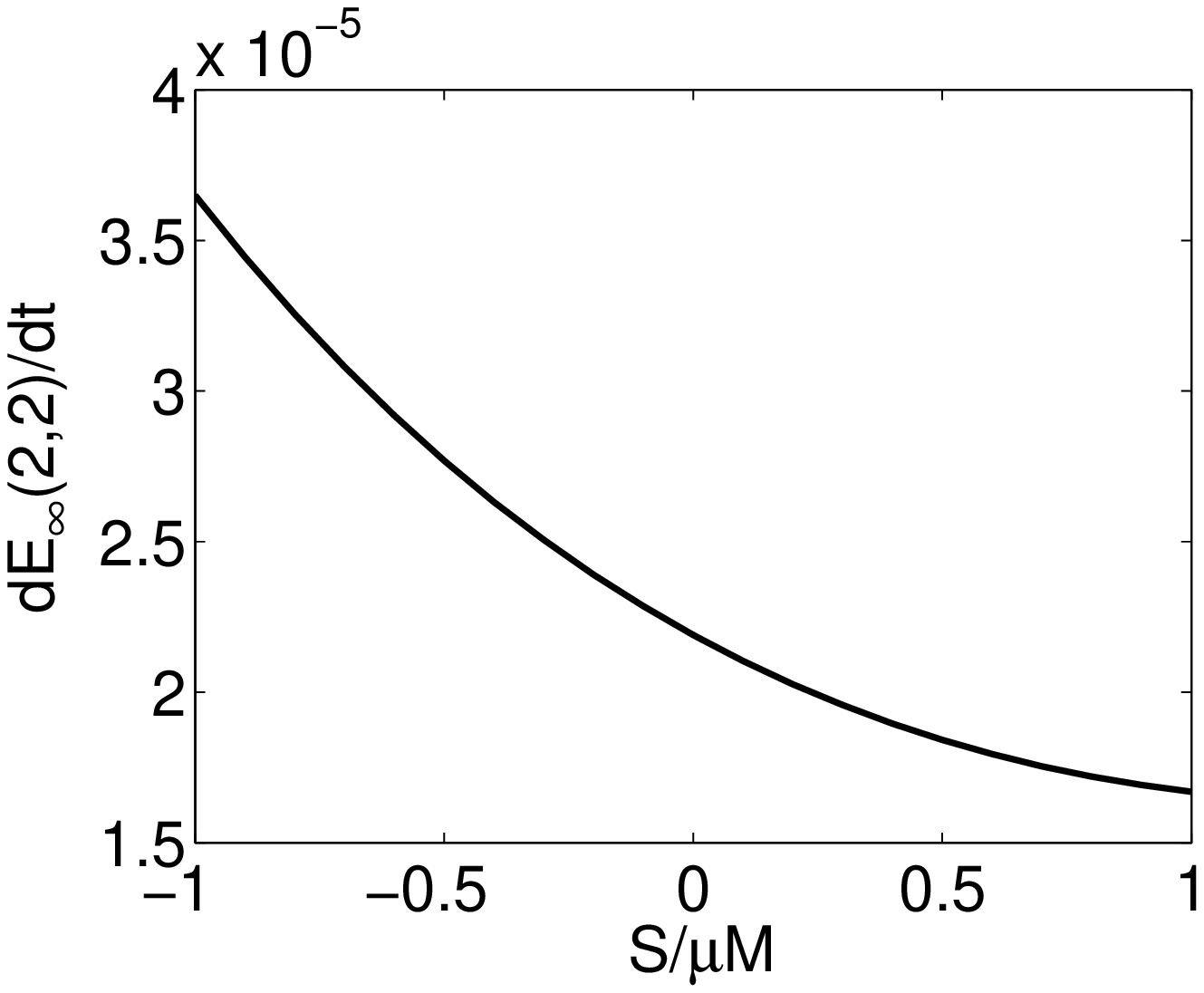}
\includegraphics[width=2.5in]{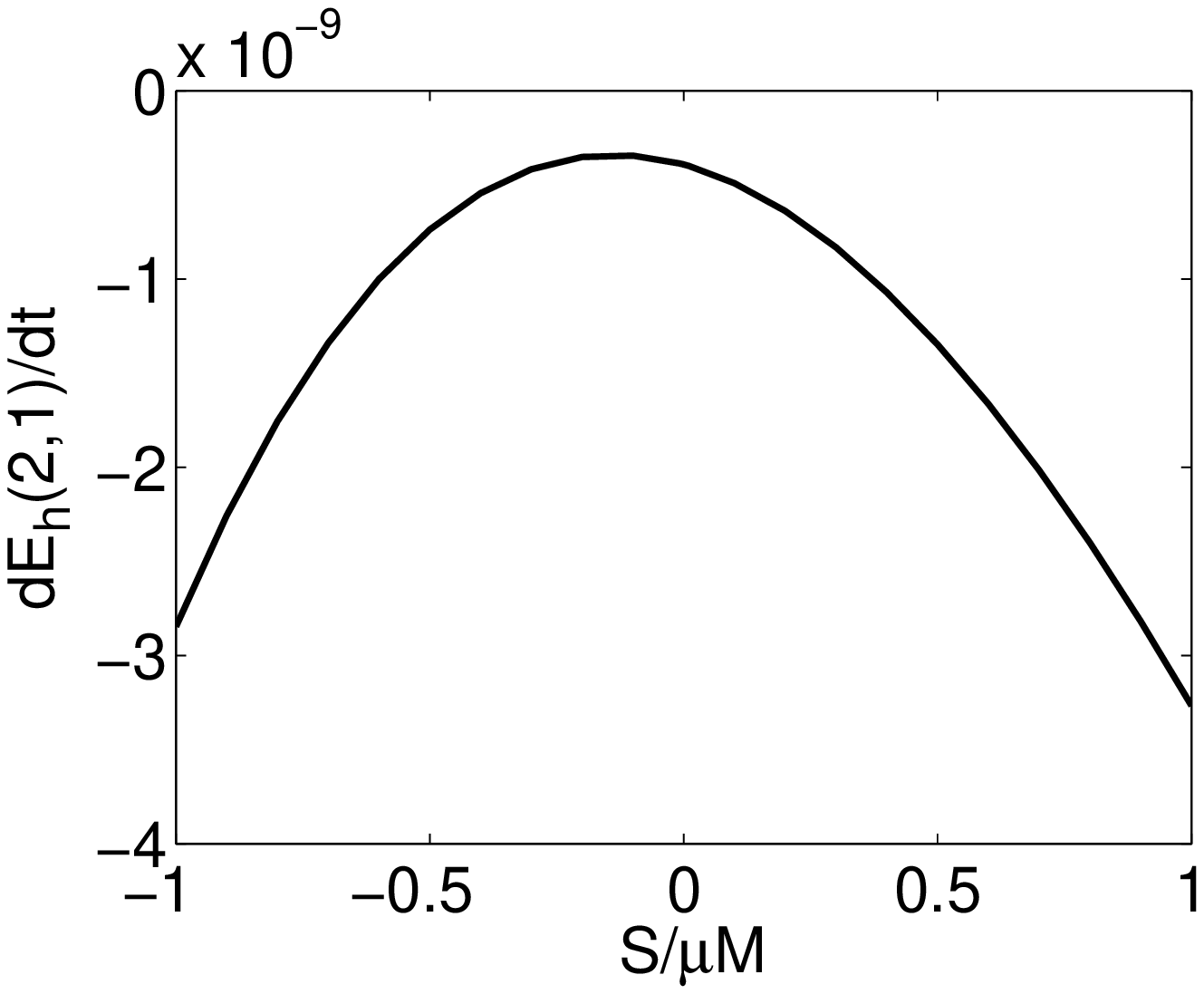}
\includegraphics[width=2.5in]{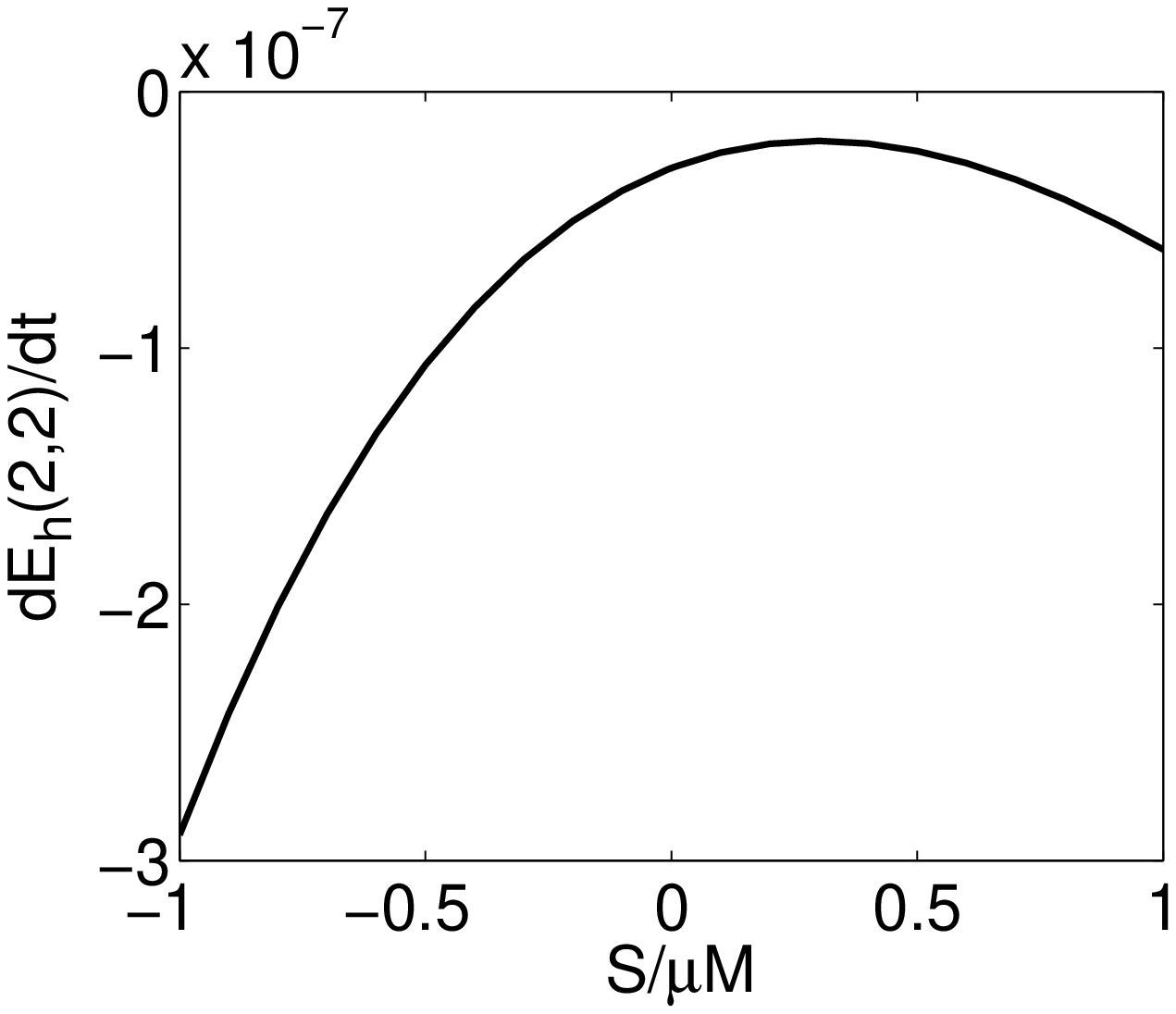}
\caption{The energy fluxes of $l=2$ modes at infinity and the
horizon as a function of the spin magnitude $S$. Where the Kerr
parameter $a=0.996$ and the orbital radius $r=10M$.}
\label{fluxlma0996}
\end{center}
\end{figure}

\begin{figure}[!h]
\begin{center}
\includegraphics[width=2.5in]{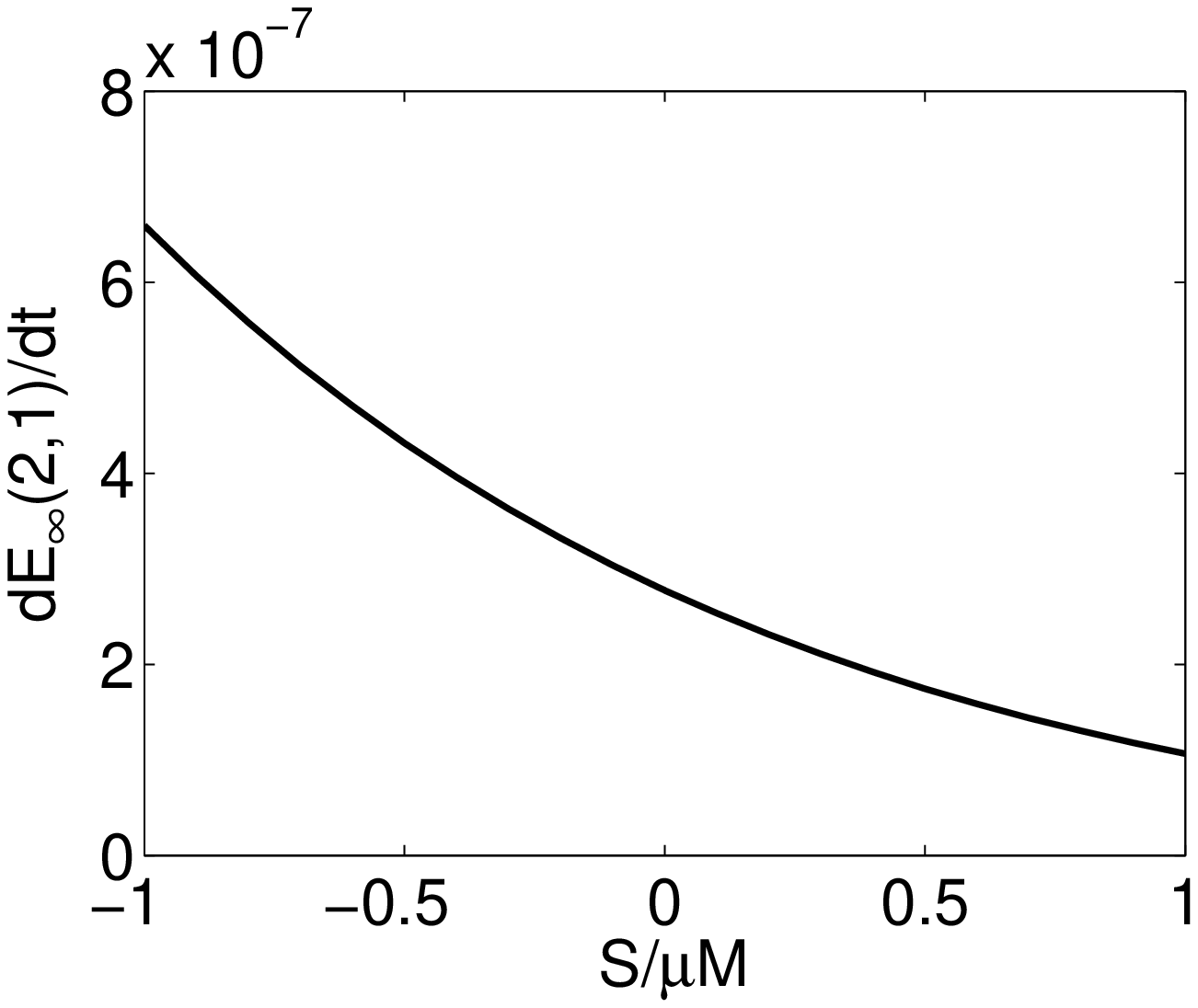}
\includegraphics[width=2.5in]{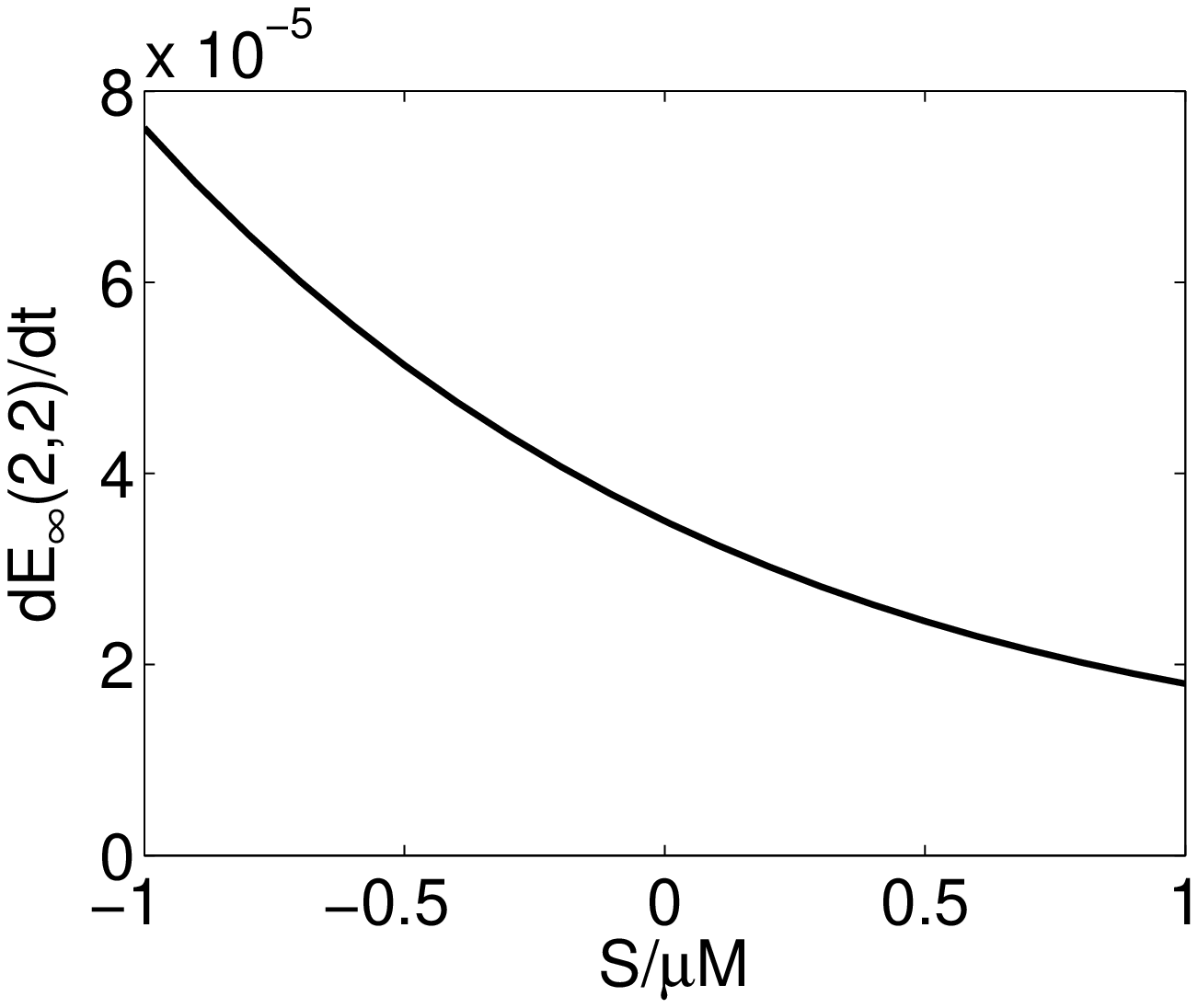}
\includegraphics[width=2.5in]{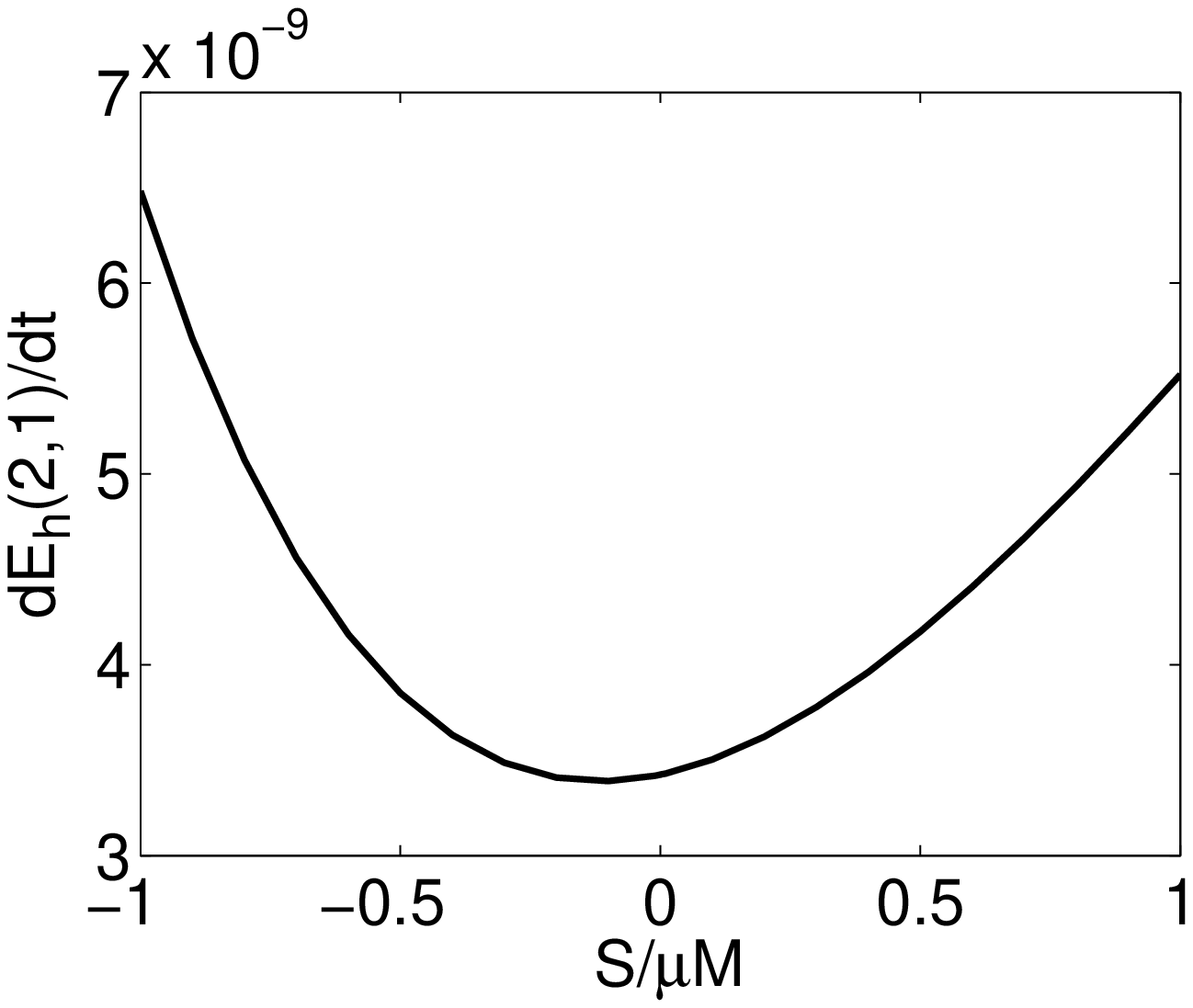}
\includegraphics[width=2.5in]{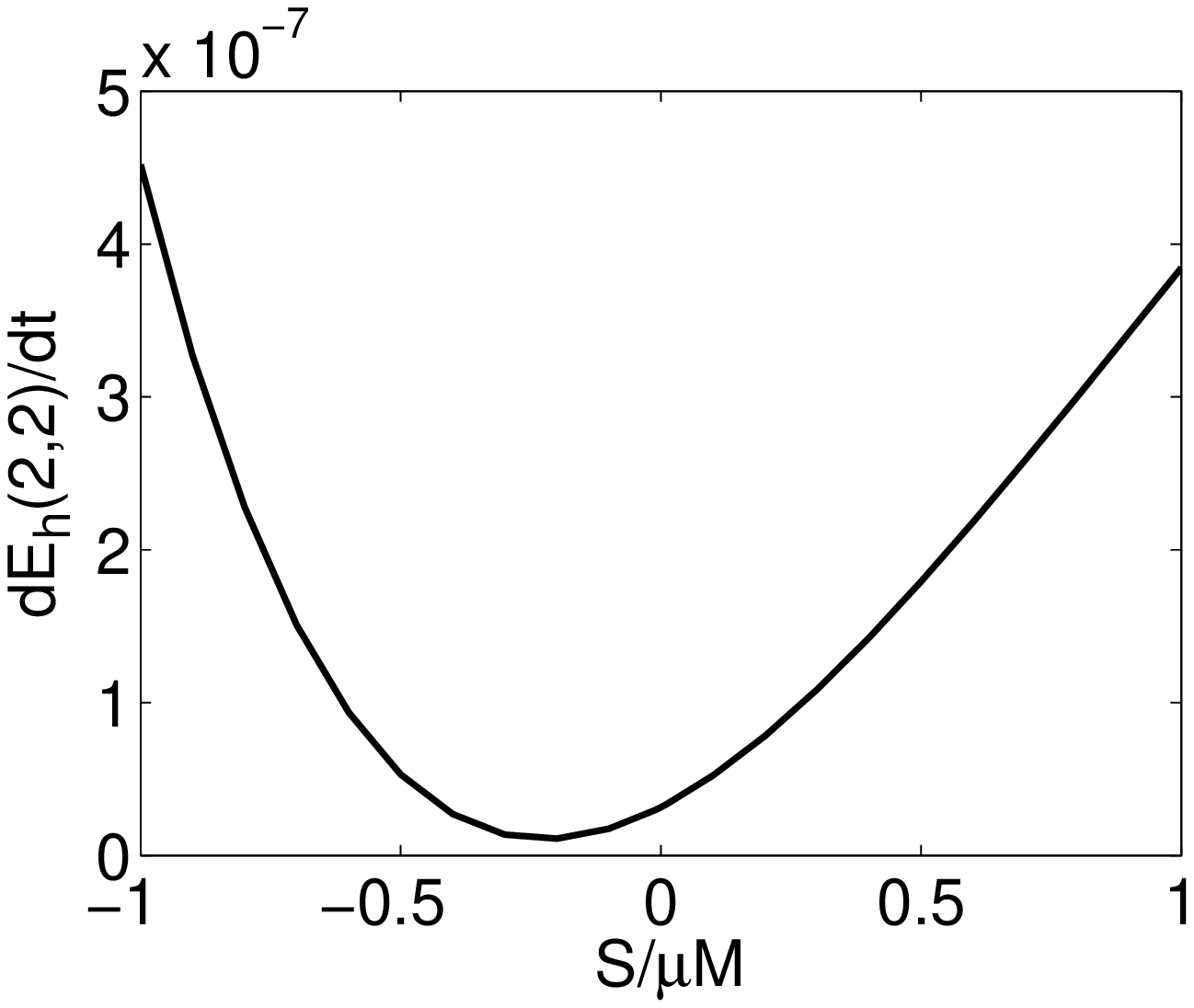}
\caption{The energy fluxes of $l=2$ modes at infinity and the
horizon as a function of the spin magnitude $S$. Where the Kerr
parameter $a=-0.996$ and the orbital radius $r=10M$.}
\label{fluxlma-0996}
\end{center}
\end{figure}

Now we show the results of energy fluxes of the spinning particles.
In Figs.(\ref{fluxlma0}, \ref{fluxlma0996} and \ref{fluxlma-0996}),
we list the energy fluxes of a particle orbiting the  black hole
with $a=0,~0.996$ and $-0.996$ respectively at infinity and the
horizon. All the orbital radiuses are $r=10$. We can find the energy
fluxes to infinity and the horizon being a function of the spin $S$
looks like quadratic functions. Usually, the $dE_\infty/dt$
decreases as the spin magnitude increases. This result coincides
with the known relation between the energy fluxes and black hole
rotation. In other words, the spin of particle together with the
black hole's rotating angular momentum, if they are ``positive''
(the same direction with orbital angular momentum), will reduce the
energy flux down to infinity, and if ``negative'', will add
$dE_\infty/dt$.
 At the same time the fluxes down to the horizon
 are no longer monotonic varying with spin; this is exhibited
 clearly in Figs.\ref{fluxlma0}-\ref{fluxlma-0996} except for the $dE_{\rm{h}}(2,1)/dt$ for $a=0$.
 This phenomenon is also analogous with
 the case of the $dE_{\rm{h}}/dt$ vs the black hole's angular momentum
 $a$. An interesting phenomenon is that when $a\leq 0$, the function
 of $dE_{\rm{h}}/dt$ of $S$ is a concave function, but a convex function when $a >
 0$.

\begin{figure}[!h]
\begin{center}
\includegraphics[width=2.5in]{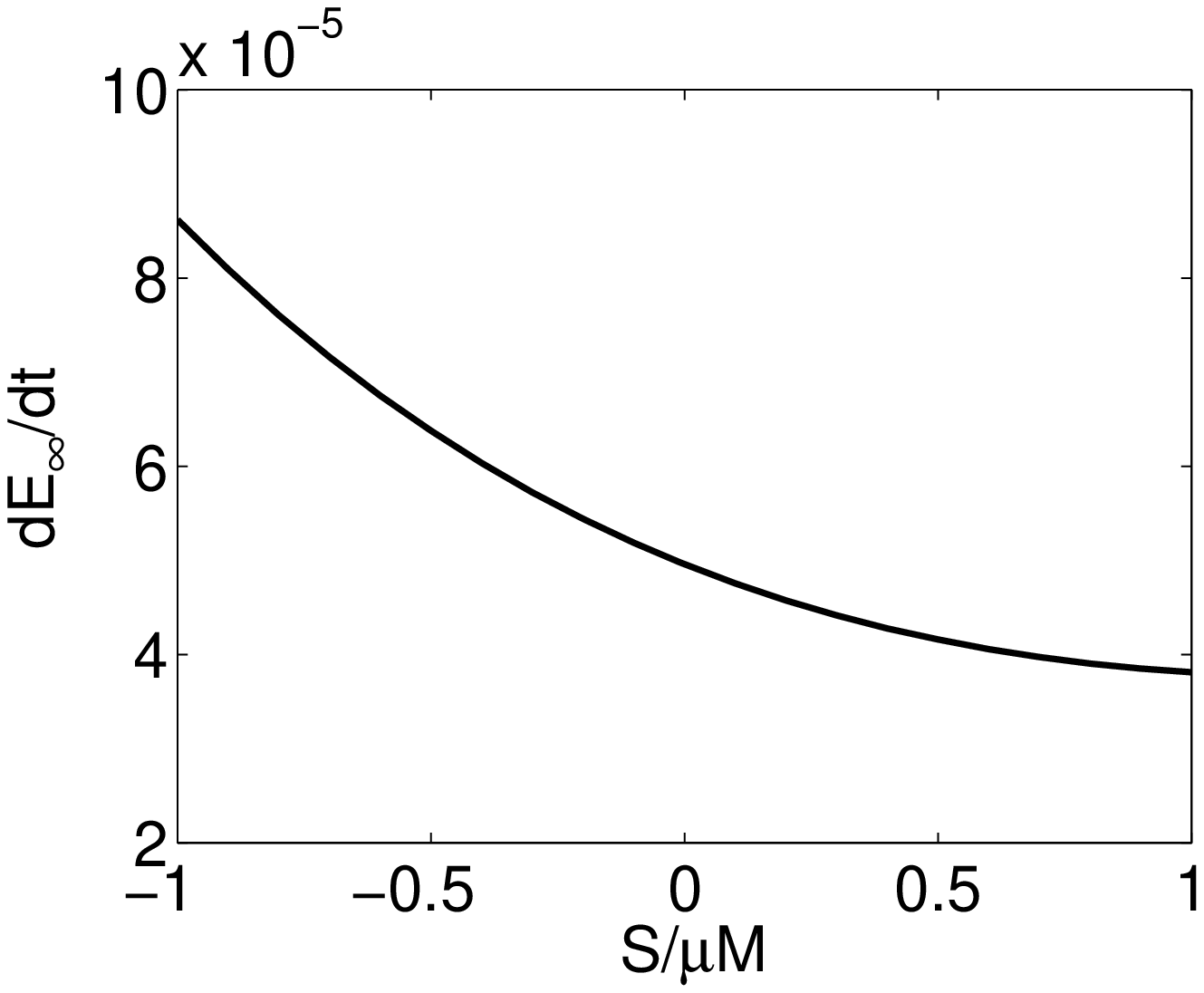}
\includegraphics[width=2.5in]{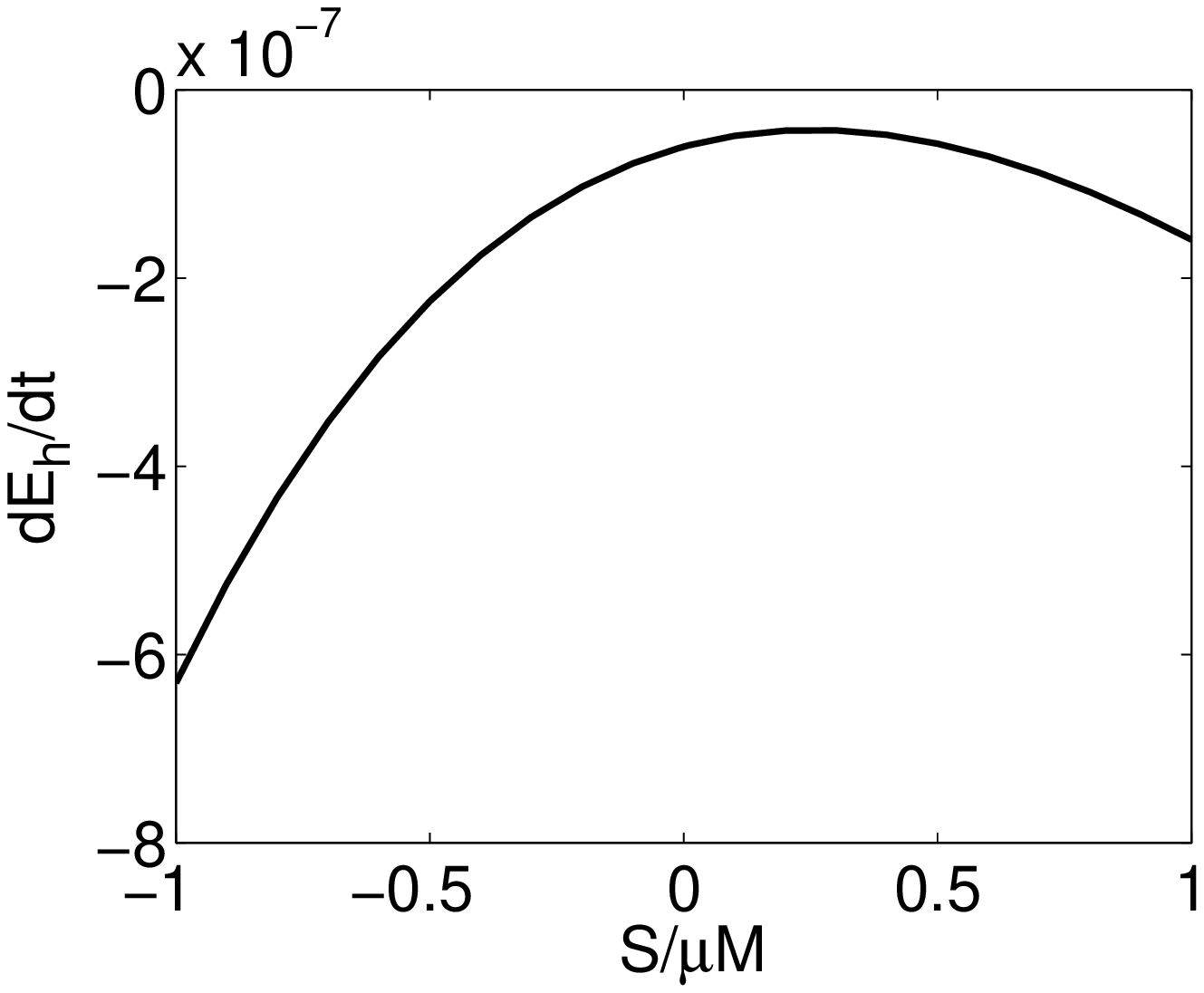}
\caption{The gravitational wave luminosity at infinity and the
horizon (up through $l=8$ to produce the accuracy) as a function of
the spin magnitude $S$. Where the Kerr parameter $a=0.996$ and the
orbital radius $r=10M$.} \label{fluxa0996}
\end{center}
\end{figure}

In Fig.\ref{fluxa0996}, we give function images of the gravitational
wave luminosity vs the spin $S$, and we produce enough accuracy of
the luminosity up through $l=8$. All of the results we figured out
should be enough to clear the effect of spin acting on the energy
fluxes. The angular momentum flux, from Eqs.(\ref{flux8},
\ref{fluxh}), is just the energy flux over $\Omega_\phi$. It is not
necessary to list the results of the angular momentum flux here.

\subsection{Energy loss vs spin}
Because of the gravitational radiation, the energy of a particle
decreases , and so does the orbital radius. The total energy loss
can be calculated by Eq.(\ref{energyloss}). Through computing the
energy-loss rates for different spin values at a certain radius, we
can compare how the spin influences the energy loss of a particle
due to gravitational radiation. For the convenience of our
comparison, we introduce relative differences of the energy-loss
rates,
\begin{align}
\frac{\Delta
\dot{E}(S)}{\dot{E}}=\frac{\dot{E}(S\neq0)-\dot{E}(S=0)}{\dot{E}(S=0)}.
\end{align}
In Table \ref{table3},  the values of $\Delta \dot{E}(S)/\dot{E}$
are listed in detail.

It is assumed that LISA will observe gravitational waves of EMRIs at
the typical frequency $\sim 10^{-2}$Hz and the total wave cycle is
about $N\sim 10^5$ for one year. Thus, the relative difference of
energy luminosity due to spin $\Delta \dot{E}(S)/\dot{E}$ required
to establish the accuracy for the cycle $\Delta N \leq 1$ must be
$\leq 10^{-5}$ in circular orbit cases \cite{Cutler}. For more
complicated orbits, such as zoom-whirl orbits, the requirement for
accuracy would be stronger than for this circular one. From Table
\ref{table3}, we can clearly find for typical $S=10^{-5}$ that the
relative differences of luminosity reach even exceed the above
requirement. The results here argue again that the spin of a small
object should be computed for the data analysis of LISA et al., and
is consistent with the previous calculation of frequency shift in
Sec. IV A.

\begin{table}
\caption{The relative differences of energy loss
$\Delta\dot{E}/\dot{E}$ vs varied spin values  for
 $a=0.996$.}\label{table3}
\begin{tabular}{c|c c c c c c}
\hline \hline
$S$&$10^{-3}$&$10^{-4}$&$10^{-5}$&$-10^{-5}$&$-10^{-4}$&$-10^{-3}$\\
\hline $r=10$&$-4.31993\times 10^{-4}$&$-4.32215\times
10^{-5}$&$-4.32303\times 10^{-6}$&$4.32140\times
10^{-6}$&$4.32255\times 10^{-5}$&$4.32475\times 10^{-4}$\\
$r=8$&$-5.98264\times 10^{-4}$&$-5.98605\times
10^{-5}$&$-5.98697\times 10^{-6}$&$5.98834\times
10^{-6}$&$5.98760\times 10^{-5}$&$5.99122\times 10^{-4}$\\
$r=6$&$-9.31789\times 10^{-4}$&$-9.32626\times
10^{-5}$&$-9.32635\times 10^{-6}$&$9.32736\times
10^{-6}$&$9.32800\times 10^{-5}$&$9.33652\times 10^{-4}$\\
$r=4$&$-1.84199\times 10^{-3}$&$-1.84453\times
10^{-4}$&$-1.84480\times 10^{-5}$&$1.84482\times
10^{-5}$&$1.84510\times 10^{-4}$&$1.84764\times 10^{-3}$\\
$r=2$&$-9.76634\times 10^{-3}$&$-9.60818\times
10^{-4}$&$-9.59232\times 10^{-5}$&$9.58882\times
10^{-5}$&$9.57298\times
10^{-4}$&$9.41434\times 10^{-3}$\\
$r=1.5$&$-7.57872\times 10^{-2}$&$-7.47114\times
10^{-3}$&$-7.46037\times 10^{-4}$&$7.45798\times
10^{-4}$&$7.44723\times
10^{-3}$&$7.33930\times 10^{-2}$\\
 \hline \hline
\end{tabular}
\end{table}

If we want to calculate the total energy radiation during the whole
inspiraling process from a big $r$ to the horizon, we need to
integrate $\dot{E}$ step by step. But this calculation requires too
much computer time, for example, for nonspinning particles, perhaps
a CPU year or more. For spinning particles on circular or equatorial
orbits, it would not add much cost [we need to calculate additional
three terms in Eq.(\ref{spinC})]. But for general orbits, it needs
to numerically work out the Papapetrou Eq.(\ref{spin1}-\ref{spin3}).
By previous experiences \cite{spin6}\cite{Hanphd}, we know that it
adds about half cpu year compared to nonspinning cases. Fortunately,
thanks to the theory which circular remains circular under adiabatic
radiation reaction, the energy loss of the particle is just the
energy difference of the particle at the circular orbit radius $r$
and near the horizon. In Table \ref{table4}, we list the particle
energy at radius $r=10M$ and $r=1.2M$ (very near the horizon), and
give the energy difference between the two cases. We can find that
the particle with positive spin will radiate more energy during the
inspiral to the black hole than nonspinning one. At the same time,
the particle with negative spin radiates less energy. Remember that
the gravitational wave luminosity decreases when spin is positive
and increases when spin is negative(see Fig.(\ref{fluxa0996})). This
means the inspiraling time of positively spinning particle should be
longer, and shorter for negative one.

Especially, for the extreme spinning particle ($S\sim1$), the
radiated energy is much more than nonspinning one (see the first row
in Table \ref{table4}). This phenomenon only occurs when $a\approx
1$ and $S$ is extreme (and the same direction as the rotation of
black hole). It is known that the maximum fractional
 binding energy per unit rest mass is $42\%$ for $a=1$ and about $30\%$ for $a=0.998$
 \cite{Hartle}. But we point out for extreme spinning particle, the
 binding energy is much larger than the nonspinning one. For
 example, for $a=0.996$ and $S=1$, the maximum fractional
 binding energy is about $77\%$. But we should take care that the
 physical value of spin is $\ll 1$. Thus, this big binding
 energy here only has theoretical meaning.

\begin{table}
\caption{The particle energy at $r=10M$, $r=1.2M$ and energy loss vs
varied spin values for $a=0.996$.}\label{table4}
\begin{tabular}{c|c c c }
\hline \hline
$S$&$E(r=10M)$&$E(r=1.2M)$&$\Delta E$\\
\hline
$1.0$&$0.9504110109413$&$0.2966990902654$&$0.6537119206759$\\
$0.1$&$0.9517720321309$&$0.6831909453624$&$0.2685810867685$\\
$10^{-3}$&$0.9519178894668$&$0.7337036132776$&$0.2182142761892$\\
$10^{-5}$&$0.9519193435776$&$0.7342411704276$&$0.2176781731500$\\
$0$&$0.9519193582782$&$0.7342466042358$&$0.2176727540423$\\
$-10^{-5}$&$0.9519193729484$&$0.7342520379715$&$0.2176673349769$\\
$-10^{-3}$&$0.9519208270598$&$0.7347903703284$&$0.2171304567313$\\
$-0.1$&$0.9520657813190$&$ 0.7932472876811$&$0.1588184936379$\\
 \hline \hline
\end{tabular}
\end{table}

\section{Conclusion and discussion}
In this paper, we have calculated the gravitational waves emitted by
a spinning particle in circular orbits on equatorial plane around a
Kerr black hole. We solved the completely orbital parameters of the
spinning particles: orbital frequency, energy and total angular
momentum and nonzero four-momenta. We also give the relation between
the four-velocity and linear momentum.

Using the Teukolsky formalism of the Kerr black hole perturbation,
we analyze the effect of spin on gravitation waves in detail. Our
calculations were done in the strong-field area and can apply to
extreme spin value. Perhaps this is the first time that the
numerical results indicate the spin of a small body cannot be
ignored for the future gravitational waveform data analysis .

The main effects of spin presented in this paper are: a positive
spin can decrease wave frequency and energy luminosity, but this
result is reversed when spin is negative; the relations of energy
flux of spin are close to quadratic functions; in the whole inspiral
process, the evolution time of positively spinning particle is
longer and the total energy radiated is larger; the overall phase
difference compared to the nonspinning case achieves the critical
value, 1 cycle; the maximum binding energy of extreme spinning
particles is much more than for a nonspinning one.

In future, it will be interesting to discuss the spin effect for
more complicated orbits such as zoom-whirl orbits with a big
eccentricity even with inclination.

\begin{acknowledgments}
The author appreciates Professor Hughes and Dr. Dolan for their
help, as well as for data provided by them to validate my results.
\end{acknowledgments}

\end{document}